\documentclass{aa}
\usepackage{graphicx}
\usepackage{natbib}
\bibpunct{(}{)}{;}{a}{}{,}

\newcommand{\Mo} {$M_{\odot}$}

\newcommand{\Hb} {H$\beta$}
\newcommand{\Ha} {H$\alpha$}
\newcommand{\nodata}{...}
\begin{document}

\title{The tidally disturbed luminous compact blue galaxy Mkn 1087 and its surroundings\thanks{Based on observations %%@
made with several telescopes operated on the islands of La Palma and Tenerife by the Isaac Newton Group of Telescopes %%@
and Instituto de Astrof\'\i sica de Canarias in the Spanish Observatories of Roque de Los Muchachos and Teide of the %%@
Instituto de Astrof\'\i sica de Canarias.}}

\author{\'Angel R. L\'opez-S\'anchez\inst{1}, C\'esar Esteban\inst{1} \and M\'onica Rodr\'{\i}guez\inst{2}}

\institute{Instituto de Astrof{\'\i}sica de Canarias, E-38200, La Laguna, Tenerife, Spain \and
Instituto Nacional de Astrof\'{\i}sica, \'Optica y Electr\'onica, Apdo. Postal 51 y 216, 72000 Puebla, Mexico}
%\email{mrodri@inaoep.mx} 

\offprints{\'Angel R. L\'opez-S\'anchez, \email{angelrls@ll.iac.es}}

\date{Received 11 June 2004 / Accepted 27 August 2004}

\abstract{We present new broad-band optical and near-infrared CCD imaging together with deep optical %%@
intermediate-resolution spectroscopy of Mkn 1087 and its surrounding objects. We analyze the morphology and colors of %%@
the stellar populations of the brightest objects, some of them star-formation areas, as well as the kinematics, %%@
physical conditions and chemical composition of the ionized gas associated with them. Mkn 1087 does not host an Active %%@
Galactic Nucleus, but it could be a Luminous Compact Blue Galaxy. Although it was classified as a suspected Wolf-Rayet %%@
galaxy, we do not detect the spectral features of these sort of massive stars. Mkn 1087 shows morphological and %%@
kinematical features that can be explained assuming that it is in interaction with two nearby galaxies: the bright KPG %%@
103a and a dwarf ($M_B\sim-18$) star-forming companion. We argue that this dwarf companion is not a tidal object but %%@
an external galaxy because of its low metallicity [12+log(O/H) = 8.24] with respect to the one derived for Mkn 1087 %%@
[12+log(O/H) = 8.57] and its kinematics. Some of the non-stellar objects surrounding Mkn 1087 are connected by bridges %%@
of matter with the main body, host star-formation events and show similar abundances despite their different angular %%@
distances. These facts, together their kinematics, suggest that they are tidal dwarf galaxies formed from material %%@
stripped from Mkn 1087. A bright star-forming region at the south of Mkn 1087 (knot \#7) does not show indications of %%@
being a tidal galaxy or the product of a merging process as suggested in previous works. We argue that Mkn 1087 and %%@
its surroundings should be considered a group of galaxies.}

\titlerunning{The tidally disturbed LCBG Mkn 1087}
\authorrunning{L\'opez-S\'anchez, Esteban \& Rodr\'{\i}guez}
\maketitle


\section{Introduction}

\begin{table*}[t!]\centering
  \caption{Summary of observations}
  \smallskip
  \label{table1}  
  \small
  \begin{tabular}{ccccccccc}
    \hline\hline
	\noalign{\smallskip}
    Observations & Telescope & Date & Exp. Time & Spatial & Filter/ & P.A. & Spectral & $\Delta\lambda$ \\
                 &           &      &    (s)  & ($\arcsec$ pix$\rm^{-1}$)& grating &($\rm^{\circ}$) & (\AA\ 
pix$\rm^{-1}$)& (\AA) \\
	\hline
	\noalign{\smallskip}
    Broad-band & 2.2m CAHA & 00/12/19 & 3   $\times$ 1200& 0.53  & $V$  & \nodata &\nodata & \nodata \\
	imaging    & 2.56m NOT  & 03/01/20 & 3   $\times$  300& 0.19  & $B$  & \nodata &\nodata & \nodata \\
	           & 2.56m NOT  & 03/01/20 & 6   $\times$  300& 0.19  & $R$  & \nodata &\nodata & \nodata \\
	           & 1.55m CST & 02/09/24 & 120 $\times$ 20&    1.0  & $J$  & \nodata &\nodata & \nodata \\
	           & 1.55m CST & 02/09/24 & 240 $\times$ 10&    1.0  & $H$  & \nodata &\nodata & \nodata \\
	           & 1.55m CST & 02/09/24 & 360 $\times$ 5 &    1.0  & $K_s$ & \nodata &\nodata & \nodata \\
    \hline
	\noalign{\smallskip}
    Intermediate & 4.2m WHT & 00/12/31& 1800 x 3 & 0.20 & R600B& 358.0 & 0.45 & 3650-5100\\
    resolution   & 4.2m WHT & 00/12/31& 1800 x 3 & 0.36 & R136R& 358.0 & 1.49 & 5300-6650\\
    spectroscopy & 2.5m INT & 99/12/28& 1200 x 4 & 0.40 & R400V& 357.0 & 1.40 & 3500-7700 \\
			     & 2.5m INT & 99/12/29& 1200 x 3 & 0.40 & R400V&   6.0 & 1.40 & 3500-7700 \\
				 & 2.5m INT & 99/12/29& 1200 x 3 & 0.40 & R400V&  62.0 & 1.40 & 3500-7700 \\
				 & 2.5m INT & 99/12/29& 1200 x 3 & 0.40 & R400V& 120.0 & 1.40 & 3500-7700 \\
				       
    \noalign{\smallskip}
	\hline\hline
  \end{tabular}
\end{table*}

One of the most controversial issues about \ion{H}{ii} galaxies is the triggering mechanism of the violent bursts of %%@
star formation that are observed in them. The estimated star formation rates in these starburst galaxies are so high %%@
that the material available for the creation of new stars would be exhausted very soon compared to the age of the %%@
universe. This problem is more marked in dwarf galaxies. \citet{Th91} suggested that the bursts have an intermittent %%@
behavior. However, star formation episodes show a large variability in duration, ranging from $10^7-10^8$ year %%@
\citep{RL85} to more than 10$^9$ yr \citep{HG85}. In fact, the age of the bursts and the galaxy that hosts them is %%@
also a controversial problem. The knowledge of both the triggering mechanism and the age of the starbursts would help %%@
to understand the evolution of the galaxies.

Hierarchical formation models of galaxies \citep{K97} predict that most galaxies have formed by
merging of small clouds of protogalactic gas. In this sense, studies of metal deficient blue compact dwarf galaxies %%@
(BCDGs) embedded in large presumably primordial \ion{H}{i} clouds provide excellent clues about the trigger and youth %%@
of the galaxies \citep{TIL95, IT99} although most of them show an underlying old stellar component \citep{N03}. Some %%@
authors \citep{SS88} proposed that galaxy interactions are the main triggering mechanism in the starbursts of dwarf %%@
and irregular galaxies. The interactions are not usual with nearby giant galaxies \citep{CA93, TT95}, but they seem to %%@
be more frequent with low surface brightness galaxies \citep{T96,WLM96}.

On the other hand, the gas-rich dwarf irregular galaxies could have formed by material ejected from the disks of the
parent galaxies to the intergalactic medium by tidal forces \citep{OT00}. The recycled objects formed in this way (the %%@
tidal dwarf galaxies, TDGs) give important clues to both age and past evolution of the galaxy \citep{DM98}. In some %%@
cases, star-forming TDGs are found at the end of long stellar tails in interacting systems \citep{WDF03}. 

In one way or in another, the interactions seem to play a key role in the triggering of the star formation in dwarf %%@
galaxies. An increasing number of starburst galaxies show Wolf-Rayet (WR) features in their integrated spectra. The %%@
so-called Wolf-Rayet galaxies have a broad emission feature around 4650 \AA\ that usually consist of a blend of lines %%@
\ion{He}{ii} $\lambda$4686 (the most important line), \ion{C}{iii}/\ion{C}{iv} $\lambda$4650 and \ion{N}{iii} %%@
$\lambda$4640, and it indicates the presence of a substantial population of this sort of massive stars. The most %%@
massive OB stars became WR stars around 2 and 3 Myr since their birth, and spend between 3 and 6 Myr in this phase %%@
\citep{MM94}. The short time in which the WR features can be observed makes these systems ideal objects to look for %%@
the probable interaction mechanism of the star formation. They also provide information about the age and properties %%@
of the starbursts. Studying a sample of WR galaxies, \citet{ME00} (hereinafter ME00) suggested that interactions with %%@
or between dwarf objects could be the main star formation triggering mechanism in dwarf galaxies. Once triggered, the %%@
evolution of a starburst depends on the details of its interaction with the galaxy environment. ME00 also remarked %%@
that the interacting and/or merging nature of the WR galaxies can only be detected when deep and high-resolution %%@
images and spectra are available.

\begin{figure*}[ht]
\includegraphics[angle=90,width=1\linewidth]{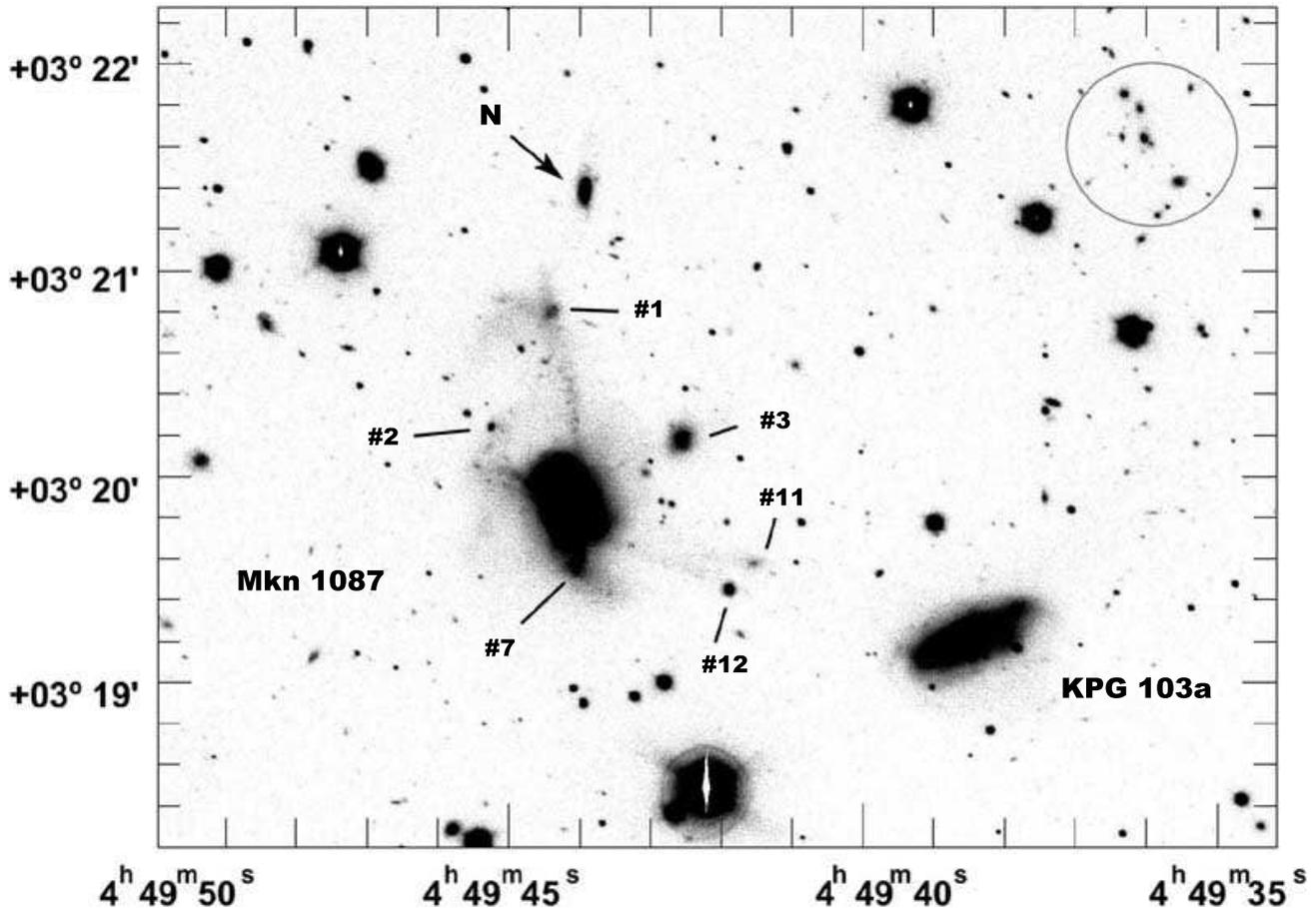}
\caption{\small{Deep optical image of Mkn 1087 and its surrounding in $R$ filter. Mkn 1087, KPG 103a, the north %%@
companion galaxy (N) and the rest of the surrounding objects are labeled. The background cluster of galaxies is %%@
indicated with a circle.}}
\label{figR}
\end{figure*}

Mkn 1087 (II Zw 23, UGC 3179) was firstly described by \citet{Zw71} as an elliptically shaped BCG with a very extended %%@
filament to the north and a less extended one to the south. These filaments were studied by \citet{K88} using optical %%@
imaging and spectroscopy and radio continuum mapping of the system. He found that the filaments were redshifted with %%@
respect to the central zones of the galaxy, suggesting that they are the result of neutral gas infalling from a very %%@
massive \ion{H}{i} cloud in which Mkn 1087 is embedded. He proposed that Mkn 1087 is a galaxy in formation and %%@
suggested that the star formation process has been occurring continuously for the past 10$^9$ year. He also remarked %%@
that it could be as a consequence of the interaction with a companion galaxy located 10$\arcsec$ to the south of the %%@
Mkn 1087 nucleus (our knot \#7, see \S3.1). Mkn 1087, which is at around 111 Mpc (assuming $H_0$= 75 km s$^{-1}$ Mpc %%@
$^{-1}$, at this distance 1$\arcsec$=538 pc), has a companion galaxy named KPG 103a, at 81 kpc to the southwest and at %%@
the same redshift \citep{M96}. \citet{MGT98} obtained a $HST$ WFPC2 image of Mkn 1087 showing that it is a spiral %%@
galaxy with dust lanes on one side of its major axis. 

The presence of WR features in the integrated spectrum of Mkn 1087 was first investigated by \citet{KJ85}, who gave an %%@
upper limit to the flux of the blue WR bump around \ion{He}{ii} $\lambda$4686. \citet{VC92} did not measure the %%@
\ion{He}{ii} $\lambda$4686 emission line but reported an upper limit of 1600 to the total late-type nitrogen-rich WR %%@
stars (WNL) contained in the central region of Mkn 1087. \citet{Va97} performed a spectroscopical study of 46 galaxies %%@
with emission line spectra in where Mkn 1087 was included, and reported a detection of the \ion{He}{ii} $\lambda$4686 %%@
emission line, giving its flux with an error less than 50\%. However, they did not give information about its line %%@
width, so the nebular or stellar origin can not be specified. As consequence, \citet{SCP99} classified Mkn 1087 as a %%@
suspected Wolf-Rayet galaxy.

ME00 presented deep CCD UBV images of Mkn 1087, finding and characterizing several faint objects connected by bridges %%@
and tails with the main body of the galaxy. They used \citet{LH95} population synthesis models to estimate the ages of %%@
the emission zones detected in their images, finding that some of these emission knots are around 5 Myr old. ME00 %%@
interpreted Mkn 1087 as a galaxy experiencing an interaction event with one or more of the neighboring objects. The %%@
main aim of this study is to perform a more complete observational analysis of these surrounding objects in order to %%@
establish their nature.

\section{Observations and data reduction}

\subsection{Optical imaging}

Optical $V$ images of Mkn 1087 were carried out on 2000 December 19 with CAFOS (Calar Alto Faint Object Spectrograph) %%@
in image mode at the Cassegrain focus of the 2.2m telescope of the Centro Astron\'omico Hispano-Alem\'an (CAHA) at %%@
Calar Alto Observatory (Almer\'{\i}a, Spain). A SITe CCD detector (2048 $\times$ 2048 pixels) with a pixel size of 24 %%@
$\mu$m and spatial resolution of 0.53$\arcsec$ pixel$^{-1}$ was used, as well as the standard Johnson $V$ filter. %%@
Three 1200 seconds exposures were added for each filter to obtain a good signal-to-noise and an appropriate 
removal of cosmic rays in the final images. The measured full width at half maximum (FWHM) of the point-spread %%@
function (PSF) was approximately 1$\arcsec$. Images were taken under photometric conditions, and the standard field %%@
98-1119 of \citet{L92} was used to perform the flux calibration.  

Images in $B$ and $R$ filters were taken on 2004 January 20 at the 2.56m Nordic Optical Telescope (NOT) at Roque de %%@
los Muchachos Observatory (La Palma, Canary Islands, Spain). We used the ALFOSC (Andalucia 
Faint Object Spectrograph and Camera) in image mode with a Loral/Lesser CCD detector (2048 $\times$ 2048 pixels) with 
a pixel size of 15 $\mu$m and spatial resolution of 0.19$\arcsec$ pixel$^{-1}$, and the standard Johnson filters. 
We obtained three 300 seconds exposures for $B$ filter and six 300 seconds exposures for $R$ filter, that were %%@
combined to obtain the final images with FWHM of the PSF of 0.8$\arcsec$. The standard star SA97-284 \citep{L92} was %%@
used to flux calibrate the final images. 

Twilight images of different zones of the sky (blank fields) were taken for each filter in order to perform the %%@
flat-field correction. All the reduction process (bias subtraction, flat-fielding and flux calibration) was done with %%@
IRAF\footnote{IRAF is distributed by NOAO which is operated by AURA Inc., under cooperative agreement with NSF} %%@
package following standard procedures. IRAF software was also used both to determine the 3$\sigma$ contours over the %%@
sky background level and to obtain the photometric values of the galaxies and the knots surrounding Mkn 1087.

We have also re-analysed the previous $U$, $B$ and $V$ images obtained by ME00 at 2.56m NOT (Nordic Optical Telescope) %%@
on 1997 February 6. See ME00 for a detailed description of these observations.  

\begin{table*}[t!]\centering
  \caption{Results of optical and near-infrared aperture photometry in Mkn 1087 and KPG 103a knots.}
  \label{table2}  
  \scriptsize
  
  \begin{tabular}{llccccccccc}
    \\
    \hline\hline
	\noalign{\smallskip}
   Knot      & E($B-V$)   &$m\rm_B$ &$m\rm_B$$^0$&$m\rm_J$$^0$& ($U-B$)$_0$&  ($B-V$)$_0$ & ($V-R$)$_0$ &                   %%@
($V-J$)$_0$  & ($J-H$)$_0$  & ($H-Ks$)$_0$ \\
    \hline
	\noalign{\smallskip}
 Mkn 1087   &  0.17$^a$ & 13.78  & 13.08$\pm$0.04  & 12.39$\pm$0.04   &$-$0.41$\pm$0.08$^b$ &   0.17$\pm$0.08  
 & 0.20$\pm$0.08    &  0.52$\pm$0.06  & 0.20$\pm$0.06   & 0.16$\pm$0.06  \\
  C  &  0.14     & 15.19  & 14.62$\pm$0.03  & 14.33$\pm$0.05   &$-$0.57$\pm$0.06$^b$ &$-$0.02$\pm$0.06  
 & 0.11$\pm$0.06    &  0.31$\pm$0.10  & 0.15$\pm$0.10   & 0.15$\pm$0.10   \\
  N    &  0.10     & 17.58  & 17.17$\pm$0.03  & 17.01$\pm$0.06   &\nodata          &$-$0.05$\pm$0.06  
 & 0.14$\pm$0.10    &  0.21$\pm$0.08  & 0.18$\pm$0.08   & 0.13$\pm$0.08  \\
  \#1   a1    &  0.07$^c$ & 19.18  & 18.89$\pm$0.05  &$>$18.7           &$-$0.75$\pm$0.15$^b$ &$-$0.01$\pm$0.08  
 & 0.10$\pm$0.08    & $<$0.2 &\nodata &\nodata\\
  \#2         &  0.07$^c$ & 20.01  & 19.72$\pm$0.05  &$>$18.7           &$-$0.78$\pm$0.15$^b$ &   0.05$\pm$0.08  
 & 0.12$\pm$0.08    & $<$1 &\nodata &\nodata\\
  \#3   b2    &  0.07$^c$ & 18.32  & 18.03$\pm$0.03  & 17.28$\pm$0.06   &   0.08$\pm$0.30$^b$ &   0.11$\pm$0.06  
 & 0.26$\pm$0.06    &  0.64$\pm$0.10  & 0.5$\pm$0.2  &\nodata\\
  \#7 W     &  0.16     & 18.29  & 17.63$\pm$0.04  &\nodata           &$-$0.87$\pm$0.08$^b$ &$-$0.04$\pm$0.08  
 & 0.03$\pm$0.08  &\nodata &\nodata &\nodata\\
  \#8 a4      &  0.08     & 19.10  & 18.77$\pm$0.05  &\nodata           &$-$0.27$\pm$0.12$^b$ &   0.07$\pm$0.10  
 & 0.17$\pm$0.08   &\nodata &\nodata &\nodata\\
  \#11        &  0.07$^c$ & 20.69    & 20.40$\pm$0.05    &\nodata   &\nodata            &   0.35$\pm$0.10  &           %%@
0.08$\pm$0.10 &\nodata&\nodata&\nodata\\
  \#12 c3     &  0.07$^c$ & 19.61  & 19.32$\pm$0.05  & 18.5$\pm$0.1     &$-$0.1$\pm$0.3   &   0.31$\pm$0.09  
 & 0.24$\pm$0.08   &  0.55$\pm$0.15 & 0.4$\pm$0.2   &\nodata\\
  \#13 a2     &  0.40     & 19.25  & 17.61$\pm$0.05  &\nodata           &$-$0.64$\pm$0.12 &   0.01$\pm$0.09  
 & 0.04$\pm$0.08   &\nodata &\nodata &\nodata \\
  \#14 c1     &  0.07     & 18.45  & 18.16$\pm$0.05  &\nodata           &$-$0.59$\pm$0.12 &   0.28$\pm$0.10  
 & 0.23$\pm$0.10   &\nodata &\nodata &\nodata\\
  \#15 d2     &  0.19     & 20.90  & 20.12$\pm$0.08  &\nodata           &$-$0.6$\pm$0.3   &   0.22$\pm$0.12  
 & 0.23$\pm$0.10   &\nodata &\nodata &\nodata\\
  \hline
  \noalign{\smallskip}
  KPG 103a    &  0.07$^c$ & 14.05    & 13.77$\pm$0.04    &\nodata   &   0.17$\pm$0.08$^b$   &   0.18$\pm$0.06  &        %%@
0.27$\pm$0.06 &\nodata&\nodata&\nodata\\ 
  K1          &  0.07$^c$ & 17.55    & 17.26$\pm$0.05    &\nodata   &$-$0.40$\pm$0.10   &   0.18$\pm$0.10  &        %%@
0.20$\pm$0.08 &\nodata&\nodata&\nodata\\
  K2          &  0.07$^c$ & 17.10    & 16.81$\pm$0.04    &\nodata   &$-$0.49$\pm$0.08   &   0.12$\pm$0.06  &       %%@
0.11$\pm$0.06 &\nodata&\nodata&\nodata\\
  K3          &  0.07$^c$ & 17.41    & 17.12$\pm$0.04    &\nodata   &$-$0.39$\pm$0.08   &   0.19$\pm$0.06  &       %%@
0.33$\pm$0.06 &\nodata&\nodata&\nodata\\  
       
    \noalign{\smallskip}
	\hline\hline
  \end{tabular}
  \begin{flushleft}
  $^a$ Average of the values obtained for C, \#7, \#8, \#13, \#14 and \#15.\\
  $^b$ Adopting the $m_U$ values from ME00.\\
  $^c$ Galactic reddening adopted from \citet{SFD98}, E$(B-V)$ = 0.07.\\
  
  \end{flushleft}
\end{table*}

\subsection{Near-Infrared imaging}

We used the 1.55m Carlos S\'anchez Telescope (CST) at Teide Observatory (Tenerife, Canary Islands, Spain) to obtain %%@
the near-infrared images on 2002 September 24. We used CAIN camera (256 $\times$ 256 pixels) with a pixel size of 40 %%@
$\mu$m and a spatial resolution of 1$\arcsec$ pixel$^{-1}$ in the wide mode to obtain images in $J$ (1.2 $\mu$m), $H$ %%@
(1.6 $\mu$m) and $K_s$ (2.18 $\mu$m) broad-band filters. We took 10 series of 6 consecutive individual 20 s exposures %%@
in $J$ filter, 10 series of 12 individual 10 s exposures in $H$ filter and 10 series of 12 exposures of 5 s duration  %%@
in $K_s$. Each series of exposures was made at slightly different positions to obtain a clean sky image. In this way, %%@
we obtained a single image of 20 minutes in $J$ and $H$ filters, and a single one of 10 minutes in $K_s$ filter. We %%@
repeated this procedure two times for $J$ and $H$ and three times for $K_s$, combining these images in a definitive %%@
one for each filter to increase the signal-to-noise ratio. The FWHM of the PSFs of these final images was %%@
2.8$\arcsec$, 2.6$\arcsec$ and 2.5$\arcsec$ for $J$, $H$ and $K_s$, respectively. Bright and dark dome flat-field %%@
exposures were taken for each filter that were combined to obtain a good flat-field image. The standard stars As01 and %%@
As11 \citep{Hu98} were used for flux calibrate the images.

\subsection{Intermediate-resolution spectroscopy}

We have obtained long-slit intermediate-resolution spectroscopy in five slit positions centered on the nucleus of the %%@
galaxy and covering most of the surrounding objects. We have used the 4.2m William Herschel Telescope (WHT) and the %%@
2.5m Isaac Newton Telescope (INT), both at Roque de los Muchachos Observatory (La Palma, Canary Islands). 

INT spectroscopy was carried out on 2000 December 28 and 29 with the IDS spectrograph at the Cassegrain focus with the %%@
235 mm camera. An EEV CCD with a configuration of 2148 $\times$ 4200 pixels of 13.5 $\mu$m was used. The slit was %%@
2.8$\arcmin$ long and 1$\arcsec$ wide. The grating R400V was used, with a dispersion of 104.5 \AA\ mm$^{-1}$ and %%@
effective spectral resolution of 1.40 \AA. Each individual spectrum covers from 3200 to 7700 \AA. The spatial %%@
resolution is 0.40$\arcsec$ pixel$^{-1}$. Four slit positions of Mkn 1087 were observed at different position angles, %%@
which were chosen in order to cover different morphological zones. Three or four 20 minutes exposures were taken and %%@
combined to obtain good signal-to-noise and an appropriate removal of cosmic rays in the final spectra for each slit %%@
position. Comparison lamp exposures of CuAr were taken after each set of spectra. The absolute flux calibration was %%@
achieved by observations of the standard stars Feige 56, Hiltner 600 and Feige 110 \citep{M88}.

An additional slit position of Mkn 1087 was obtained using 4.2m WHT. It was carried out on 2000 December 31 with the %%@
ISIS spectrograph at the Cassegrain focus. Two different CCDs were used at the blue and red arms of the spectrograph: %%@
an EEV CCD with a configuration of 4096 $\times$ 2048 pixels of 13 $\mu$m in the blue arm and a TEK with 1024 $\times$ %%@
1024 of 24 $\mu$m in the red arm. The dichroic used to separate the blue and red beams was set at 5400 \AA. The slit %%@
was 3.7$\arcmin$ long and 1$\arcsec$ wide. Two gratings were used, the R600B in the blue arm and the R316R in the red %%@
arm. These gratings give reciprocal dispersion of 33 and 66 \AA\ mm$^{-1}$, and effective spectral resolutions of 2.0 %%@
and 3.9 \AA\ for the blue and red arms, respectively. The blue spectra cover from 3600 to 5200 \AA\ and the red ones %%@
from 5400 to 6800 \AA. The spatial resolution is 0.20$\arcsec$ pixel$^{-1}$ in the blue and 0.35$\arcsec$ pixel$^{-1}$ %%@
in the red. We obtained three 30 minutes exposures in each arm that were combined to obtain the final blue and red %%@
spectra for this slit position. Comparison lamp exposures of CuAr for the blue arm and CuNe for the red one were taken %%@
after each set of spectra. The absolute flux calibration was achieved by observations of the standard stars G191, B2B %%@
and Feige 34 \citep{M88}. WHT observations were made at air masses around 1.1. INT observations were made at air %%@
masses between 1.1 and 1.3. Consequently, no correction was made for atmospheric differential refraction.   

All the CCD frames for spectroscopy were reduced using standard IRAF procedures to perform bias correction, %%@
flat-fielding, cosmic-ray rejection, wavelength and flux calibration, and sky subtraction. The correction for %%@
atmospheric extinction was performed using an average curve for the continuous atmospheric extinction at Roque de los %%@
Muchachos Observatory. For each two-dimensional spectra several apertures were defined along the spatial direction to %%@
extract the final one-dimensional spectra of each knot, centering each aperture at its brightest point.
For WHT observations, identical apertures were defined on the "blue" and "red" frames. Small two-dimensional %%@
distortions were corrected fitting the maxima of [\ion{O}{ii}] $\lambda\lambda$3726,3729 doublet and H$\alpha$ in INT %%@
frames, using also H$\beta$ and continuum emission for WHT ones. 

The journal of all the imaging and spectroscopical observations can be found in Table~\ref{table1}.

\section{Results}

\subsection{Optical and near-infrared imaging}

The deep CAHA $V$ image of the system was presented in \citet{ELSM03}, showing the main non-stellar objects %%@
surrounding Mkn 1087 found by ME00. We designated the objects following those authors. Several bridges can be observed %%@
connecting the main body of Mkn 1087 with the objects \#1, \#2, \#3, \#11 and \#12, the last two ones located in the %%@
direction of the nearby galaxy KPG 103a. Knot \#7 seems to be an intense star-formation zone off-center Mkn 1087. 

\begin{figure}[dh!]
\includegraphics[width=1\linewidth]{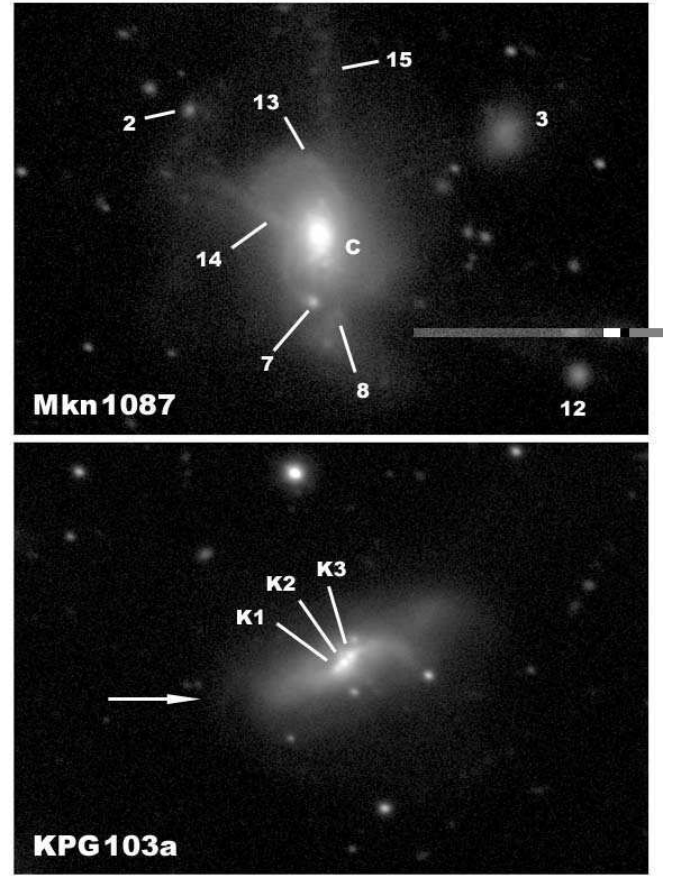}
\caption{\small{Detailed images in $R$ band of Mkn 1087 (up) and KPG 103a (bottom). The brightest knots inside both %%@
galaxies are labeled, as well as the external objects \#2, \#3 and \#12. A weak plume at the east of KPG 103a, that %%@
curves toward the south, is also barely observed (it is indicated by an arrow).}}
\label{fig3}
\end{figure}

In order to obtain a deeper image, we carried out new observations of the system in $R$ broad-band filter. We present %%@
this image in Figure~\ref{figR}. The bridges commented before are now clearly detected, especially the ones that end %%@
at \#11 and \#12. The bright bridge connecting Mkn 1087 with \#1 is apparently splitted in several small knots. The %%@
north companion object seems to be clearly elliptical and shows a very weak plume in the direction to knot \#1. %%@
Although bridges between the two brightest galaxies are not detected, a long very weak plume can be observed at the %%@
east of KPG 103a. Another diffuse long plume is detected at the west. It is important to note the high number of %%@
(apparently) background non-stellar objects (presumably galaxies) that are detected in the new deep $R$ image (we have %%@
estimated around 60 candidates in the entire final $R$ frame). They are uniformly distributed in the field of view of %%@
the image (about 6$\arcmin$) and are most probably background galaxies. We also want to remark the cluster of galaxies %%@
found at the upper right corner of the image, with central position around $\alpha$ = 04h 49min 35sec, $\delta$ = %%@
+03$^\circ$ 21$\arcmin$ 04$\arcsec$. It is not catalogued in the NASA/IPAC Extragalactic Database (NED). We analyze it %%@
in \S4.7.       

\begin{figure*}[t]
\includegraphics[width=1\linewidth]{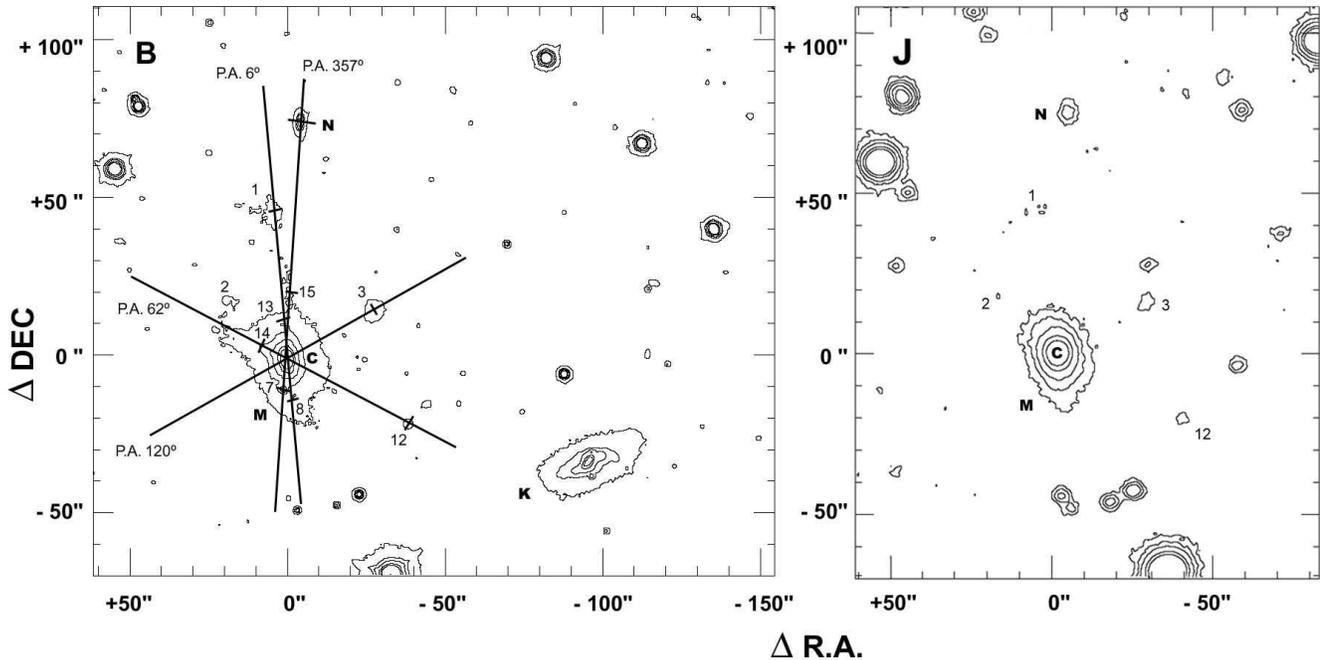}
\caption{\small{Logarithmic contour maps in $B$ and $J$ filters of Mkn 1087. M indicates the main body of Mkn 1087, C %%@
the center of it, K the nearby galaxy KPG 103a and N the north companion galaxy. The contours are 3$\sigma$, %%@
10$\sigma$, 20$\sigma$, 50$\sigma$ and 100$\sigma$ over the sky level for both maps. The slit positions used for %%@
spectroscopy are also shown over the $B$ image, indicating with ticks the center of the different zones from which we %%@
have extracted the one-dimensional spectra.}}
\label{rendijas}
\end{figure*}

Some important structures inside the main galaxies are now clearly distinguished. In Figure~\ref{fig3} we show the %%@
central part of Mkn 1087 (up) and KPG 103a (down). The center of Mkn 1087, C (knot \#5 following ME00), is the %%@
brightest knot of the system. Mkn 1087 reveals two spiral arms or structures: an upper arm connecting with the bridge %%@
towards knot \#2 and a bottom arm that seems to end in knot \#7. Furthermore, some new knots are now detected inside %%@
the main body of both galaxies. Especially intense are the three bright knots K1, K2 and K3, that are resolved in the %%@
center of KPG 103a. They show blue colors (see Table~\ref{table2}) that are representative of young population, %%@
suggesting that these knots are star-formation areas in the nucleus of KPG 103a. The WFPC2 $HST$ image of Mkn 1087 %%@
presented by \citet{MGT98} clearly shows knots C, \#7, \#8, \#13 and \#14 and the arms of the spiral disk we have %%@
distinguished in our NOT images. The $HST$ image suggests that object \#2 could be a background object that is %%@
coincident with the eastern bridge of Mkn 1087, although this cannot be confirmed without a spectrum of the object. %%@
However, this knot shows very blue colors, $(U-B) = -0.78$, similar to other knots around Mkn 1087, so we prefer to %%@
consider knot \#2 in our study.

In Figure~\ref{rendijas} we show the logarithmic contour map of Mkn 1087 and its surroundings in filters $B$ and $J$. %%@
We clearly detect the north companion object in all NIR images. However, the knots \#3 and \#12 are only detected in %%@
$J$ and $H$ filters. Knots \#1 and \#2 are barely detected in the $J$ image. We did not observe the galaxy KPG 103a in %%@
NIR.  

ME00 performed aperture photometry of the knots of the system in $U$, $B$ and $V$ filters. We have improved their %%@
analysis performing the aperture photometry of the companion object in our new $B$, $V$ and $R$ images. We have %%@
determined the area for which we have integrated the flux using the 3$\sigma$ level isophote over the average sky %%@
level in the $B$ image. The same integration area was then used for $V$ and $R$ filters. We have also measured the %%@
rest of the knots in the system, finding similar values to those obtained by ME00. In addition, we have determined the %%@
photometric $U$, $B$ and $V$ values of some new interesting areas (\#13, \#14 and \#15) because they correspond to %%@
localized bright nebula emission in the disk of Mkn 1087 as our spectroscopy reveals (see next section), they probably %%@
correspond to giant \ion{H}{ii} regions. Finally, we have analyzed the NIR images to obtain the photometric $J$, $H$ %%@
and $K_s$ values of the detected objects. The integration area in $J$ filter (defined as the 3$\sigma$ level isophote %%@
over the sky level) was used for each NIR filter. We also estimated a lower limit for $m_J$ in knots \#1 and \#2. All %%@
these new optical and NIR data are shown in Table~\ref{table2}. We only give upper limits to the $J$ magnitudes for %%@
knots \#1 and \#2. \citet{HG85} performed $JHK$ photometry within 23$\arcsec$ aperture and found $J-H$ = 0.65, $H-K$ = %%@
0.28 and $V-J$ = 1.76, suggesting a dominant contribution of old populations at these wavelengths. 

\begin{table*}[t]\centering
 \caption{General spectroscopic properties of the knots}
 \label{table3}
 \scriptsize
 \begin{tabular}{lcccccccccccc}
   \\
   \hline\hline
   \noalign{\smallskip}
     Knot & C WHT & C INT & N WHT & N INT & 7 WHT &\#7 INT& \#1   & \#3    & \#8     &\#13    &\#14   &\#15  \\
   \hline
    \noalign{\smallskip}
$-$M$_B$  & 20.61 & 20.61 & 18.06 & 18.06 & 17.60 & 17.60 & 16.34 & 17.20  &  16.46  & 17.62  & 17.07 & 15.11 \\
 $\pm$    &  0.03 &  0.03 &  0.03 &  0.03 &  0.04 &  0.04 &  0.05 &  0.03  &   0.05  &  0.05  &  0.05 &  0.08 \\
 \noalign{\smallskip}
 \noalign{\smallskip}
Slit ($\arcsec\times$1$\arcsec$)&5.4&12&3.6&8.4&6.0&  3.6 &  5.6  &   4.0  &   5.2   &  7.2   &  5.6  & 8.0   \\
Distance ($\arcsec$)$\rm^a$&0&0&77.0& 77.0 &  9.8  & 10.0 & 47.6  &  35.2  &  10.9   &  10.8  &  8.0  & 19.2  \\
 \noalign{\smallskip}
 \noalign{\smallskip}
C(H$\beta$)$\rm^b$&0.17& 0.25&0.17&  0.12 &  0.24 &  0.25 &  0.10 &  0.10  &   0.12  &  0.62  &  0.11 &  0.29  \\
 $\pm$    &  0.02 &  0.03 &  0.03 &  0.03 &  0.03 &  0.03 &  0.03 &  0.03  &   0.03  &  0.12  &  0.04 &  0.05  \\
 \noalign{\smallskip}
$W_{abs}$$\rm^{b,c}$&1.7&2.4& 0.2 &  0.9  &  2.9  &  1.5  &\nodata&\nodata &   1.1   &   2    &  5.6  &   2    \\
 $\pm$    &   0.1 &   0.3 &  0.2  &  0.2  &  0.5  &  0.3  &\nodata&\nodata &   0.3   &\nodata &  1.7  &\nodata \\
 \noalign{\smallskip}
 \noalign{\smallskip}
$T_e[$O III$]\rm^d$&0.71&0.74& 1.09& 1.13 & 0.77  &  0.78 & 0.82  &  0.76  &  0.88   &   0.82 &  0.86 &  0.81  \\
$T_e[$O II$]\rm^e$& 0.80& 0.82&1.06& 1.09 & 0.84  &  0.84 & 0.88  &  0.83  &  0.92   &   0.87 &  0.90 &  0.87  \\
$N_e$ (cm$^{-3}$)&220& 170 &  115  &  150 & 120   &   210 &\nodata&\nodata &  150    &   120  & $<$100&$<$100  \\
 \noalign{\smallskip}
 \noalign{\smallskip}
$\Delta$v$_r$$\rm^f$&0 & 0 & +117  & +110  & +118  & +122  & $-$63 &   +8   &  +125   & $-$85  & $-$70 &$-$75  \\
 $\pm$  &\nodata  &\nodata&   15   &  27   &  15   &  27   &  27   &  80    &  27     &  27    &  35   &  35   \\
 \noalign{\smallskip}
 \noalign{\smallskip}
$F$(H$\beta$)$\rm^g$&133& 184& 4.58& 11.3  & 16.6  & 1.76  & 3.05: & 1.58:  & 6.03    & 1.92   & 8.46  & 0.94: \\
 $\pm$              & 4 &  6 & 0.16&  0.6  &  0.7  & 0.09  &\nodata&\nodata & 0.37    & 0.26   & 0.62  &\nodata \\
\noalign{\smallskip}
\noalign{\smallskip}
$W$(H$\alpha$)$\rm^c$& 87.4 & 87.3 & 53.6 & 66.5  & 330  & 256   & 110   & 20:   & 72.0   & 30.9   & 140  & 39.7  \\
$\pm$                &  2.6 &  3.4 &  2.1 &  8.2  &  25  &  28   &  25   &\nodata&  4.5   &  2.8   &  20  &  9.5  \\
\noalign{\smallskip}
$W$(H$\beta$)$\rm^c$ & 22.3 & 22.9 & 25.0 & 27.4  & 62.3 & 66    &\nodata&\nodata& 16.6   & 5.6    & 52   & 8.0:  \\
$\pm$                &  0.9 &  2.1 &  1.6 &  2.4  &  5.4 & 12    &\nodata&\nodata&  2.6   & 1.3    &  8   &\nodata\\
\noalign{\smallskip}
$W$(H$\gamma$)$\rm^c$&  8.1 &  6.6 & 8.2  &  5.3  &34.3  & 20.2  &\nodata&\nodata&  8.3   &\nodata & 18   &\nodata\\
$\pm$                &  0.4 &  0.6 & 0.9  &  0.5  & 4.9  &  2.2  &\nodata&\nodata&  1.6   &\nodata &  5   &\nodata\\
\noalign{\smallskip}
$W$([O III])$\rm^c$  & 16.7 & 20.5 & 144  & 46.6  & 116  & 63.2  &\nodata&\nodata& 25.7   & 6.8    & 42.0 & 7.8:  \\
$\pm$                &  0.6 &  0.9 &  18  &  3.7  &  9   &  3.6  &\nodata&\nodata&  2.8   & 1.2    &  3.6 &\nodata\\

\noalign{\smallskip}
\noalign{\smallskip}
Age$\rm^h$           & 6.0  & 6.0  &  7   &  6    & 5.0  &  4.5  &  6     & 150  &  6     & 8.5    &    5  &  8.0  \\
 $\pm$               & 0.5  & 0.5  &  1   &  1    & 0.5  &  0.5  &  2     &  30  &  1     & 1.5    &    1  &  1.5  \\           %%@
\noalign{\smallskip}
 \hline\hline
 \end{tabular}
 \begin{flushleft}
 $\rm^a$ Distance from the center of Mkn 1087. \\
 $\rm^b$ C(H$\beta$) and $W_{abs}$ calculated iteratively. See \S 3.2 for more details.\\
 $\rm^c$ In \AA.\\
 $\rm^d$ In units of $10^4$K, calculated using empirical calibrations of \citet{P01}\\
 $\rm^e$ In units of $10^4$K, $T_e [$OII$]$ calculated from empirical calibration and the relation given by %%@
\citet{G92}\\
 $\rm^f$ Radial velocity with respect to the center of Mkn 1087, in km s$^{-1}$  \\
 $\rm^g$ In units of 10$\rm^{-16}$ erg s$\rm^{-1}$ cm$\rm^{-2}$ \\
 $\rm^h$ Estimated ages using our spectroscopical and photometric data, in Myr. See \S 4.1 for more details.\\

 \end{flushleft}
\end{table*} 

The reddening constant, C(H$\beta$) ---obtained from the absorption-corrected intensities of 
\ion{H}{i} Balmer lines in our optical spectra of each member--- was used to correct the photometric data for the %%@
extinction by interstellar dust. We describe the method used to obtain C(H$\beta$) from the spectra in the next %%@
section. The relation obtained by \citet{KL85} between C(H$\beta$) and the extinction in $V$ filter ($A_V$) was used %%@
[$A_V\sim$ 2C(H$\beta$)], and a standard ratio of $R_V$ = $A_V$/$E(B-V)$ = 3.1 was assumed. \citet{C00} propose a %%@
starburst reddening curve with $R_{V} = 4.05 \pm 0.80$. If we use this curve for our objects, the $B$ magnitudes would %%@
be between 0.08 and 0.24 brighter than the ones we have calculated assuming $R_{V} = 3.1$. We would also obtain bluer %%@
colors, but the differences are smaller than the photometric uncertainties. The choice between the two reddening %%@
curves implies only slight changes in the results and  does  not change our conclusions. The data in $U$, $B$, $R$, %%@
$J$, $H$ and $K_s$ filters were corrected using the \citet{RL85} extinction correction law for $A_U$, $A_B$, $A_R$, %%@
$A_J$, $A_H$ and $A_{Ks}$. For the main body of Mkn 1087 we have assumed an average extinction using the objects C, %%@
\#7, \#8, \#13, \#14 and \#15, that are the knots inside Mkn 1087 for which we have spectra. \citet{SFD98} presented %%@
an all-sky map of infrared dust emission, determining a color excess of $E(B-V)$ = 0.07 (assuming $R_V$ = 3.1) at the %%@
Galactic longitude and latitude of Mkn 1087. This value should be considered as a lower limit to the real one. We have %%@
assumed only this Galactic contribution to the reddening in the direction of Mkn 1087 for the more external objects %%@
(\#1, \#2, \#3 and \#12) and the galaxy KPG 103a because we have no determination of C(H$\beta$) for it. The final %%@
$E(B-V)$ adopted for each individual object is also shown in Table~\ref{table2}. We have determined the errors for the %%@
photometry considering the FWHM of the PSF, sky level and flux calibration for each frame. The error in $V-J$ color %%@
due to the differences in the shape and size between the optical and the NIR integrated areas is small when compared %%@
with the other uncertainties.

\subsection{Intermediate resolution spectra}

\begin{figure*}[th]
\centering
\includegraphics[angle=270,width=1\linewidth]{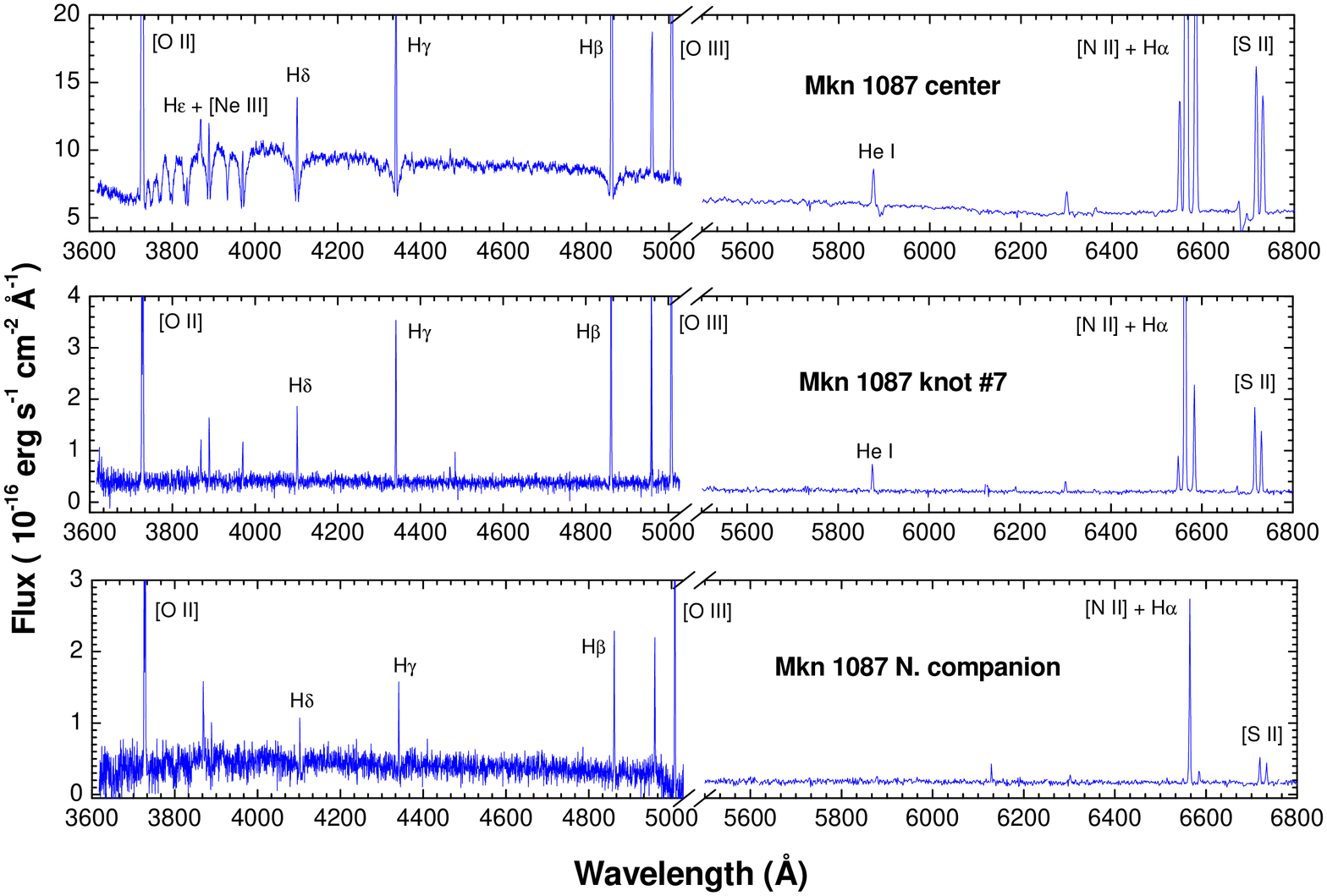}
\protect\caption[ ]{\small{ISIS WHT spectra of the center of Mkn 1087, the intense knot \#7 and its north companion %%@
object. The most important emission lines are labeled in the upper spectrum. All the spectra have been scaled down in %%@
flux in order to distinguish the faint lines. Note the high continuum level and the \ion{H}{i} absorptions due to the %%@
underlying stellar population in the spectrum of the center of Mkn 1087. All spectra are redshift-corrected.}}
\label{fig6}
\end{figure*}

\begin{table*}\centering
  \caption{Dereddened line intensity ratios with respect to I(H$\beta$)=100 for the three brightest star-forming %%@
bursts observed in Mkn 1087: the center of Mkn 1087 (C), the north companion galaxy (N) and the intense burst \#7. All %%@
have been observed both with 4.2m WHT and 2.5m INT. Correction for underlying stellar absorption in \ion{H}{i} Balmer %%@
lines has been also applied.}
  \label{table4a}  
  \footnotesize
  \begin{tabular}{lrcccccc}
  \noalign{\smallskip}
    \hline\hline
	\noalign{\smallskip}
	Line  & f($\lambda$)  &   C WHT      &   C INT      &  N  WHT    &   N  INT     &  \#7 WHT    &  \#7 INT     \\
	\hline    
	\noalign{\smallskip}
 3726 $[$O II$]$  & 0.27  &206$\pm$11$^a$&226$\pm$13$^a$&155$\pm$12  &325$\pm$26$^a$&85.5$\pm$7.1 &212$\pm$20$^a$\\
 3729 $[$O II$]$  & 0.27  & \nodata      & \nodata      &231$\pm$14  &  \nodata     &106$\pm$8    &  \nodata     \\
 3869 $[$Ne III$]$& 0.23  &8.3$\pm$2.9   &8.7$\pm$3.3   &43.4$\pm$6.9&37.8$\pm$9.1  &9.9$\pm$3.6  &17.1$\pm$6.2  \\ 
 3889 He I+H8	  & 0.22  &13.8$\pm$4.3  &7.6$\pm$2.6   &\nodata     &\nodata       &15.0$\pm$4.3 &15.6$\pm$4.6  \\
 3968 $[$Ne III$]$& 0.21  &9.7$\pm$3.5   &3.54:         &\nodata     &\nodata       &11.0$\pm$3.6 &13.2$\pm$3.7  \\
 4068 $[$S II$]$  & 0.19  &2.23:         &\nodata       &\nodata     &\nodata       &\nodata      &\nodata       \\
 4076 $[$S II$]$  & 0.19  &1.13:         &\nodata       &\nodata     &\nodata       &\nodata      &\nodata       \\
 4101 H$\delta$  & 0.18   &25.9$\pm$4.3  &25.9$\pm$4.6  &25.9$\pm$4.8&25.9$\pm$7.7  &25.9$\pm$4.4 &25.9$\pm$4.7  \\ 
 4340 H$\gamma$  & 0.135  &46.9$\pm$4.7  &46.9$\pm$5.2  &46.9$\pm$7.2&46.9$\pm$8.8  &46.9$\pm$8.1 &46.9$\pm$9.8  \\
 4363 $[$O III$]$ & 0.13  &1.76:         &\nodata       &\nodata     &\nodata       & \nodata     &1.85:         \\
 4471 He I	     & 0.10   &2.14:         &3.90$\pm$1.5  &\nodata     &\nodata       &3.23:        &\nodata       \\
 4861 H$\beta$   & 0.00   &100$\pm$5     &100$\pm$6     &100$\pm$5   &100$\pm$7     &100$\pm$5    &100$\pm$6     \\
 4959 $[$O III$]$&$-$0.02 &32.2$\pm$2.3  &33.9$\pm$2.7  &78.2$\pm$5.4&124$\pm$7     &69.0$\pm$5.5 &57.7$\pm$6.1  \\
 5007 $[$O III$]$&$-$0.03 &97.4$\pm$5.9  &109$\pm$7     &218$\pm$10  &375$\pm$19    &201$\pm$12   &161$\pm$10    \\
 5016 He I	    &$-$0.03  &1.18:         &\nodata       &\nodata     &\nodata       &\nodata      &\nodata       \\
 5200 $[$N I$]$  &$-$0.05 &\nodata       &2.97$\pm$0.79 &\nodata     &\nodata       &\nodata      &\nodata       \\
 5518 $[$Cl III$]$&$-$0.17&\nodata       &0.94:         &\nodata     &\nodata       &\nodata      &\nodata       \\
 5876 He I	    &$-$0.23  &10.3$\pm$1.2  &11.1$\pm$1.9  &12.6$\pm$2.2&14.5$\pm$2.4  &10.3$\pm$1.3 &12.6$\pm$1.9  \\
 6300 $[$O I$]$  &$-$0.30 &6.73$\pm$0.60 &7.56$\pm$0.76 &19.1$\pm$3.0&9.7$\pm$2.2   &4.55$\pm$0.79&4.9$\pm$1.9   \\
 6364 $[$O I$]$  &$-$0.31 &1.83:         &1.77:         &\nodata     &\nodata       &1.41:        &\nodata       \\
 6548 $[$N II$]$ &$-$0.34 &31.4$\pm$2.7  &31.5$\pm$3.3  &8.0$\pm$1.7 &9.4$\pm$3.3   &14.2$\pm$1.9 &14.9$\pm$2.2  \\ 
 6563 H$\alpha$ &$-$0.34  &286$\pm$11    &286$\pm$12    &286$\pm$12  &286$\pm$14    &286$\pm$12   &286$\pm$14    \\
 6584 $[$N II$]$ &$-$0.34 &102.2$\pm$6.4 &98.1$\pm$8.8  &22.6$\pm$2.7&22.8$\pm$4.1  &43.1$\pm$3.0 &46.5$\pm$5.4  \\ 
 6678 He I	    &$-$0.35  &1.84:         &\nodata       &\nodata     &10.3$\pm$2.9  &2.09:        &\nodata       \\
 6716 $[$S II$]$ &$-$0.36 &42.9$\pm$3.8  &43.8$\pm$3.9  &48.7$\pm$4.9&34.8$\pm$4.3  &31.0$\pm$3.4 &32.0$\pm$5.7  \\
 6731 $[$S II$]$ &$-$0.36 &36.0$\pm$2.9  &35.1$\pm$4.2  &37.4$\pm$4.1&27.4$\pm$3.6  &23.7$\pm$2.2 &26.4$\pm$2.9  \\
 \noalign{\smallskip}
    \hline\hline
  \end{tabular}
   \begin{flushleft}
   $^a$ $[$O II$]$ $\lambda\lambda$3726,3729 emission flux.\\
   \end{flushleft}
\end{table*}

\begin{table*}\centering
  \caption{Dereddened line intensity ratios with respect to I(H$\beta$)=100 for the faint knots observed in Mkn 1087. %%@
Correction for underlying stellar absorption in \ion{H}{i} Balmer lines was also applied. All spectra were taken using %%@
INT.}
  \label{table4b}  
  \footnotesize
  \begin{tabular}{lrcccccc}
  \noalign{\smallskip}
    \hline\hline
	\noalign{\smallskip}
	Line  & f($\lambda$)  &  \#1      &  \#3      &  \#8         & \#13       & \#14        & \#15    \\
	\hline    
	\noalign{\smallskip}
 3727 $[$O II$]$  & 0.27  & 265:      & 240:      &290$\pm$35   &252$\pm$75   &271$\pm$26   &286.39:    \\
 3889 He I+H 8	 & 0.22   &\nodata    &\nodata    &19.5$\pm$6.4 &\nodata      &\nodata      &\nodata    \\ 
 4340 H$\gamma$  & 0.135  &\nodata    &\nodata    &46.9$\pm$7.9 &\nodata      &47$\pm$14    &\nodata    \\
 4861 H$\beta$   & 0.00   & 100:      &100:       &100$\pm$9    &100$\pm$23   &100$\pm$12   &100:       \\
 4959 $[$O III$]$&$-$0.02 &\nodata    &\nodata    &56.7$\pm$8.5 &55$\pm$14    &50.7$\pm$6.6 &34.7:      \\
 5007 $[$O III$]$&$-$0.03 & 136:      & 99:       &165$\pm$14   &144$\pm$22   &173$\pm$14   &86.5:      \\
 5876 HeI	    &$-$0.23  &\nodata    &\nodata    &\nodata      &\nodata      &11.4$\pm$4.1 &\nodata    \\
 6548 $[$N II$]$ &$-$0.34 &\nodata    &\nodata    &16.8$\pm$5.9 &26$\pm$10    &13.0$\pm$4.5 &\nodata    \\ 
 6563 H$\alpha$ &$-$0.34  &286$\pm$46 &286$\pm$54 &286$\pm$23   &286$\pm$31   &286$\pm$10   &286$\pm$34 \\
 6584 $[$N II$]$ &$-$0.34 &54$\pm$18  & 51.5:     &60.0$\pm$9.1 &51.0$\pm$7.1 &59.7$\pm$7.2 &71$\pm$25  \\ 
 6716 $[$S II$]$ &$-$0.36 &\nodata    &\nodata    &45.7$\pm$7.4 &46.7$\pm$7.6 &43.2$\pm$6.5 &22:        \\
 6731 $[$S II$]$ &$-$0.36 &\nodata    &\nodata    &36.0$\pm$7.9 &36.0$\pm$5.8 &27.4$\pm$5.4 &57:        \\
 \noalign{\smallskip}
    \hline\hline
  \end{tabular}
   \begin{flushleft}
   \end{flushleft}
\end{table*}

\begin{table*}[th]\centering
 \caption{Chemical abundances of the knots.}
 \label{table4s}
 \scriptsize
 \begin{tabular}{lcccccccccccc}
 \noalign{\smallskip}
   \hline\hline
   \noalign{\smallskip}
    Knot           & C WHT & C INT& N WHT & N INT  &\#7 WHT&\#7 INT& \#1   & \#3    & \#8   &\#13   &\#14   &\#15 \\
    \hline
    \noalign{\smallskip}
12+log O/H$^a$     & 8.57  & 8.53 & 8.23   & 8.24  & 8.55  & 8.53  & 8.50: & 8.52:  & 8.41  & 8.47  & 8.43  & 8.44  \\
12+log O/H$^b$     & 8.79  & 8.78 & 8.31   & 8.32  & 8.52  & 8.54  & 8.59  & 8.58   & 8.63  & 8.57  & 8.62  & 8.68  \\
(O$^{++}$+O$^+$)/O$^+$&1.55& 1.53  & 1.53  & 1.99  & 2.19  & 1.85  &\nodata&\nodata & 1.57  & 1.64  & 1.61  & 1.67  \\
12+log N/H         & 7.76  & 7.71  & 6.77  & 6.90  & 7.48  & 7.43  &\nodata&\nodata & 7.33  & 7.48  & 7.32  & 7.32  \\
$-$ log N$^+$/O$^+$& 0.81  & 0.83  & 1.46  & 1.34  & 1.08  & 1.10  &\nodata&\nodata & 1.08  & 0.99  & 1.11  & 1.00  \\
12+log Ne$^{++}$/H$^+$& 7.67& 7.60 & 7.53  & 7.42  & 7.56  & 7.79  &\nodata&\nodata &\nodata&\nodata&\nodata&\nodata\\
12+log S$^+$/H$^+$ & 6.54  & 6.50  & 6.24  & 6.07  & 6.31  & 6.34  &\nodata&\nodata & 6.38  & 6.44  & 6.33  & 6.45  \\
12+log S$^{++}$/H$^+$ & 6.42  &\nodata&\nodata&\nodata&\nodata&\nodata&\nodata&\nodata %%@
&\nodata&\nodata&\nodata&\nodata\\
12+log He$^+$/H$^+$$^c$&10.84 & 10.88 & 10.97 & 11.03 & 10.82 & 10.94 &\nodata&\nodata &\nodata&\nodata& 10.91 %%@
&\nodata\\
 \hline
 \noalign{\smallskip}
[O/H]/[O/H]$_{\odot}$$^d$&0.76&0.69&0.35& 0.35& 0.72  & 0.69  & 0.72: & 0.65:  & 0.60  & 0.71  & 0.55  & 0.56  \\

 \hline
 \hline
 \end{tabular}
 \begin{flushleft}
 $^a$ Determined using the empirical calibration of \citet{P01}\\
 $^b$ Determined using the $[$N II$]$/H$\alpha$ ratio \citep{D02}\\
 $^c$ Using only the He I $\lambda$5875 emission line.\\
 $^d$ Asuming 12+$\log$[O/H]$_\odot$ = 8.69$\pm0.05$ \citep{AP01}\\
 \end{flushleft}

\end{table*} 

We have observed five slit positions in Mkn 1087 in order to obtain the spectra of the most interesting zones. In %%@
Figure~\ref{rendijas} we show the four different position angles (P.A.) obtained using the INT over the $B$  contour %%@
map, as well as the center of each extracted 1-D spectrum. All four slit positions cover the central zone, C, of Mkn %%@
1087 and were combined to obtain an individual high signal-to-noise spectrum of this zone. The slit with P.A. %%@
6$^\circ$ covers knots \#1, \#8 and the new bright emission area \#13. The slit with P.A. 62$^\circ$ passes through %%@
knots \#12 (that does not show nebular emission) and the new object \#14. Knot \#3 was observed with P.A. 120$^\circ$, %%@
but it only reveals a faint nebular emission. The new companion galaxy and the intense knot \#7 were observed using %%@
the slit position with P.A. 357$^\circ$, that also shows the new nebular zone \#15. The single slit position observed %%@
using WHT has a P.A. 358$^\circ$ and is not represented in Figure~\ref{rendijas} (it is similar to the P.A. %%@
357$^\circ$ shown in the figure) and covers the center of Mkn 1087, the companion galaxy and knot \#7. The distance to %%@
the center of Mkn 1087 and the size of each extracted 1-D spectrum is indicated in Table~\ref{table3}. Note that we %%@
have observed six different areas at different radii along the main body of Mkn 1087, two of them (C and \#7) observed %%@
both with INT and WHT. 

The wavelength and flux calibrated spectra of the center of Mkn 1087, the north companion and knot \#7 are shown in %%@
Figure~\ref{fig6}. The spectrum of the center of Mkn 1087 shows a high continuum level and an intense underlying %%@
stellar absorption in \ion{H}{i} Balmer lines. Absorption wings are \textbf{clearly} resolved for \Hb, H$\gamma$ and %%@
H$\delta$. The high continuum level and these absorption features can be interpreted as the product of a substantial %%@
population of stars with ages larger than 10 Myr \textbf{(see \S4.1)}. The spectra of the other regions or knots show %%@
an almost \textbf{absent} continuum level with no evident absorption features, characteristic of a \ion{H}{ii} %%@
region-dominated zone. 

Fluxes and equivalent widths of the emission lines for each spectrum were measured using the SPLOT routine of the IRAF %%@
package. This task integrates all the flux in the line between two given limits and over a fitted local continuum. We %%@
also used Starlink DIPSO \citep{HM90} software for the Gaussian fitting of profiles affected by line-blending (mainly, %%@
the [\ion{O}{ii}] $\lambda\lambda$3726,3729 doublet in WHT spectra) and to obtain the best values for the fluxes of %%@
\ion{H}{i} Balmer lines. For each single or multiple Gaussian fit, DIPSO gives the fit parameter (radial velocity %%@
centroid, Gaussian sigma, FWHM, etc) and their associated statistical errors. To correct the observed line intensities %%@
for interstellar reddening we have used the reddening constant, C(H$\beta$), derived from the fluxes of \ion{H}{i} %%@
Balmer lines. However, we must also take into account the effect of the underlying stellar absorption in the hydrogen %%@
lines. Their equivalent widths, $W_{abs}$, are usually considered, but their values are uncertain because they also %%@
depend on the age of burst \citep{O95}. We have performed an iterative procedure to derive both, C(\Hb) and $W_{abs}$, %%@
for each spectrum, assuming that the equivalent width of the absorption lines is the same for all the Balmer lines and %%@
using the relation given by \citet{MB93} to the absorption correction together with the \citet{W58} law:
\begin{eqnarray}
C(H\beta)=\frac{1}{f(\lambda)} \log\Bigg[\frac{\frac{I(\lambda)}{I(H\beta)}\times %%@
\Big(1+\frac{W_{abs}}{W_{H\beta}}\Big)} {\frac{F(\lambda)}{F(H\beta)}\times %%@
\Big(1+\frac{W_{abs}}{W_{\lambda}}\Big)}\Bigg]
\end{eqnarray}
for each detected hydrogen Balmer line. In this equation, $F(\lambda)$ and $I(\lambda)$ are the observed and the %%@
theoretical fluxes (not affected by reddening or absorption), whereas $W_{abs},\ W_{\lambda}$ and $W_{H\beta}$ are the %%@
equivalent widths of the underlying stellar absorption, the considered Balmer line and H$\beta$, respectively. %%@
$f(\lambda)$ is the reddening curve normalized to \Hb. We have considered the theoretical ratios between each pair of %%@
\ion{H}{i} Balmer lines expected for case B recombination given by \citet{B71} and appropriated electron temperatures %%@
between 6000 and 10000 K and electron densities around 100 cm$^{-3}$. The C(\Hb) and $W_{abs}$ that provide the best %%@
match between the corrected and the theoretical line ratios for each knot are shown in Table~\ref{table3}. We have %%@
assumed $W_{abs}$ = 2 \AA\ in knots \#13 and \#15 because we only detected the two brightest \ion{H}{i} Balmer lines. %%@
For \#1 and \#3 we have not corrected for W$_{abs}$ due to their faintness absorption. 

Finally, all the emission lines detected were corrected for reddening using the \citet{W58} law and the corresponding %%@
value of C(H$\beta$). The H$\beta$ line flux, $F(H\beta)$ (corrected by reddening and underlying stellar absorption) %%@
and the equivalent width of several lines (H$\alpha$, H$\beta$, H$\gamma$ and [\ion{O}{iii}]) are given in %%@
Table~\ref{table3}, while the reddening-corrected line intensity ratios relative to H$\beta$ of each observed object %%@
are given in Tables~\ref{table4a} and \ref{table4b}. We have estimated the uncertainties in the line intensities %%@
following the equation given by \citet{Ca00}. Colons indicate uncertainties of the order or greater than 40\%.

\subsubsection{Physical conditions of the ionized gas}

The 1-D spectra were used to study the physical conditions and chemical abundances of the ioni\-zed gas. %%@
Unfortunately, the auroral [\ion{O}{iii}] $\lambda$4363 line is not detected or well measured, so we can not derive %%@
the electron temperature of the ionized gas. Although [\ion{O}{iii}] $\lambda$4363 seems to be detected in the %%@
spectrum of the center of Mkn 1087, the broad absorption wing in H$\gamma$ does not permit a proper measurement of %%@
this weak emission line. Consequently, the oxygen abundance for the different knots in Mkn 1087 was derived using the %%@
so-called empirical calibrations (see next section). The electron temperatures have been estimated from the %%@
T([\ion{O}{iii}]) and T([\ion{O}{ii}]) pairs that reproduce the total oxygen abundance obtained applying the %%@
\citet{P01} empirical method (see next section). As Pilyugin, we have assumed a two-zone scheme, the usual relation %%@
O/H = O$^+$/H$^+$ + O$^{++}$/H$^+$ and the linear relation between T([\ion{O}{iii}]) and T([\ion{O}{ii}]) based on %%@
photoionization models obtained by \citet{G92}:
\begin{eqnarray}
T([\textsc{O~ii}]) = 0.7 \cdot T([\textsc{O~iii}]) + 3000. 
\end{eqnarray}
The empirical relation between the oxygen abundance and the T([\ion{O}{iii}]) given by \citet{P01} (his equation 11) %%@
gives similar T[\ion{O}{iii}] results than the ones obtained before. The finally adopted T[\ion{O}{iii}] and %%@
T[\ion{O}{ii}] for each burst are shown in Table~\ref{table3}. Following \citet{P01}, we could assume an uncertainty %%@
between 500 and 1000 K for T[\ion{O}{iii}].

We have used the [\ion{S}{ii}] $\lambda\lambda$ 6717,6731 doublet to determine the electron density, $N_e$, of the %%@
ioni\-zed gas. Electron densities were calculated making use of the five-level program for the analysis of %%@
emission-line nebulae that is included in IRAF NEBULAR task \citep{SD95}. For zones \#14 and \#15, electron densities %%@
are below the low-density limit ($<$100 cm$^{-3}$). We have no determination of $N_e$ for \#1 and \#3. The final %%@
adopted values of $T_e$ and $N_e$ (or their upper limits) are compiled in Table~\ref{table3}. 

\begin{figure*}[ht!]
\centering
\includegraphics[angle=270,width=1\linewidth]{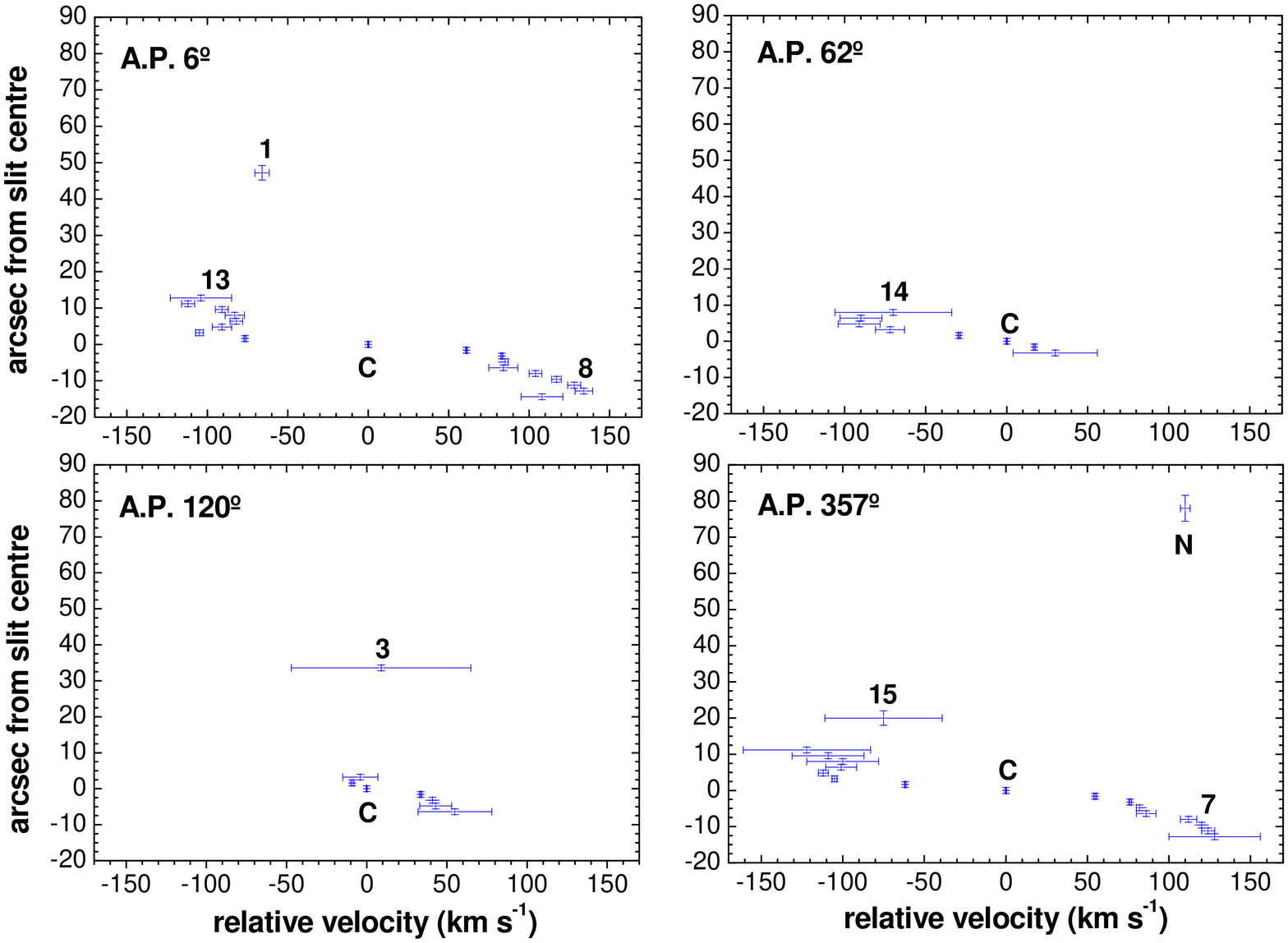}
\protect\caption[ ]{\small{a, b, c and d: Position-velocity diagrams for the four slit positions observed with INT,  %%@
analysed in 1.6 arcsec bins. The horizontal bars represent the uncertainty of the Gaussian fitting for each point. The %%@
location and extension of the different galaxy members are also indicated. North is up in all the diagrams.}}
\label{pv1}
\end{figure*}

\subsubsection{Abundance Analysis}

We have derived the total oxygen abundance making use of empirical calibrations, based on the relative intensities of %%@
strong optical lines. \citet{D02} used the [\ion{N}{ii}] $\lambda$6583 /\Ha\ ratio in combination with photoionization %%@
models to obtain the calibration:
\begin{eqnarray}
12+\log\frac{\rm{O}}{\rm{H}}=9.2+0.73\log\frac{[\textsc{N~ii}]~6583}{H\alpha}. 
\end{eqnarray}
However, the most widely used empirical abundance calibrator is the $R_{23}$ parameter, defined as \citep{Pa80}:
\begin{eqnarray}
R_{23}\equiv\frac{I([\textsc{O~ii}]\ 3727)+I([\textsc{O~iii}]\ 4959 + 5007)}{I(H\beta)}.
\end{eqnarray}
\citet{P00} found that the previous calibrations using this parameter had a systematic error depending on the hardness %%@
of the ionizing radiation, suggesting that the excitation parameter, $P$, is a good indicator of it. This parameter %%@
was defined as:
\begin{eqnarray}
P\equiv\frac{I([\textsc{O~iii}]~4959 + 5007)}{I(H\beta)}\frac{1}{\rm{R_{23}}}.
\end{eqnarray}
\citet{P00, P01} performed a detailed analysis of the observational data combined with photoionization models to %%@
obtain empirical calibrations for the oxygen abundance. \citet{P00} proposes a linear fit that involves only the %%@
$R_{23}$ parameter (for 12+log(O/H)$>$ 8.15), whereas \citet{P01} is the following, more real and complex, calibration %%@
involving also the excitation parameter:
\begin{eqnarray}
12+\log\frac{\rm{O}}{\rm{H}}=\frac{R_{23}+54.2+59.45P+7.31P^2}{6.01+6.71P+0.371P^2+0.243R_{23}},
\end{eqnarray}
(the so-called \emph{P-method}), that can be used in moderately high-metallicity \ion{H}{ii} regions. 

In Table~\ref{table4s} we show the oxygen abundances obtained for each observed knot using both empirical %%@
calibrations.  We have assumed an error of $\sim\, \pm 0.2$ dex for the oxygen abundances using \citet{D02}. %%@
\citet{P01} estimates an uncertainty of $\pm 0.10$ dex for his calibration. We find that the O/H ratios obtained using %%@
the calibration of \citet{D02} are systematically slightly higher that those obtained using the calibrations of %%@
\citet{P01}. Only in the bright burst \#7 both calibrations give a similar abundance. \citet{D02} remarked that their %%@
calibration has a scatter that seems to depend on the degree of ionization. In consequence, we have adopted the oxygen %%@
abundances derived using the P-method as the more appropiates ones. 

Note that the north companion object, N, has a clearly lower oxygen abundance [around 8.24, in units of 12+log (O/H)] %%@
compared with the rest of the knots (between 8.41 and 8.57). \citet{VC92} reported an oxygen abundance of 8.55 for the %%@
main body of Mkn 1087 assuming $T_e$= 10000 K and $N_e$= 340 cm$^{-3}$, value that coincides with the abundance we %%@
obtain for the central zone of the galaxy. The average value among all the observed knots inside Mkn 1087 is %%@
12+log(O/H) = 8.49.

In Table~\ref{table4s} we list the ionic and O abundances we have derived. All the ionic abundances have been %%@
calculated using the IRAF NEBULAR task from the intensity of collisionally excited lines, assuming the %%@
T([\ion{O}{iii}]) for the high ionization potential ions O$^{++}$, Ne$^{++}$, and S$^{++}$, and T([\ion{O}{ii}]) for %%@
the low ionization potential ions O$^+$, N$^+$ and S$^+$. 

We have assumed the standard ionization correction factor (icf) by \citet{PC69}, N/O = N$^+$/O$^+$, to derive the %%@
nitrogen abundance of the objects. This correction factor is listed in Table~\ref{table4s}. Except for the central %%@
zone of Mkn 1087, only the S$^+$/H$^+$ ratio is derived, so we are not able to derive the total sulphur abundance. %%@
Furthermore, some contribution of S$^{3+}$ is also expected in the areas with a higher ionization degree.

The He$^+$/H$^+$ ratio has been derived from the \ion{He}{i} $\lambda$5875 line observed in each knot and using the %%@
predicted line emissivities calculated by \citet{SSM96}. We have also corrected for collisional 
contribution following the calculations by \citet{B02}. In Table~\ref{table4s} we 
present the He$^+$/H$^+$ ratios finally obtained. The results obtained from INT and WHT data for a given object are %%@
quite similar, despite the uncertainties. This indicates the good quality of the spectra and the reduction and %%@
analysis procedures.   

\subsubsection{Kinematics of Mkn 1087}

\begin{figure}[t!]
\centering
\includegraphics[angle=270,width=1\linewidth]{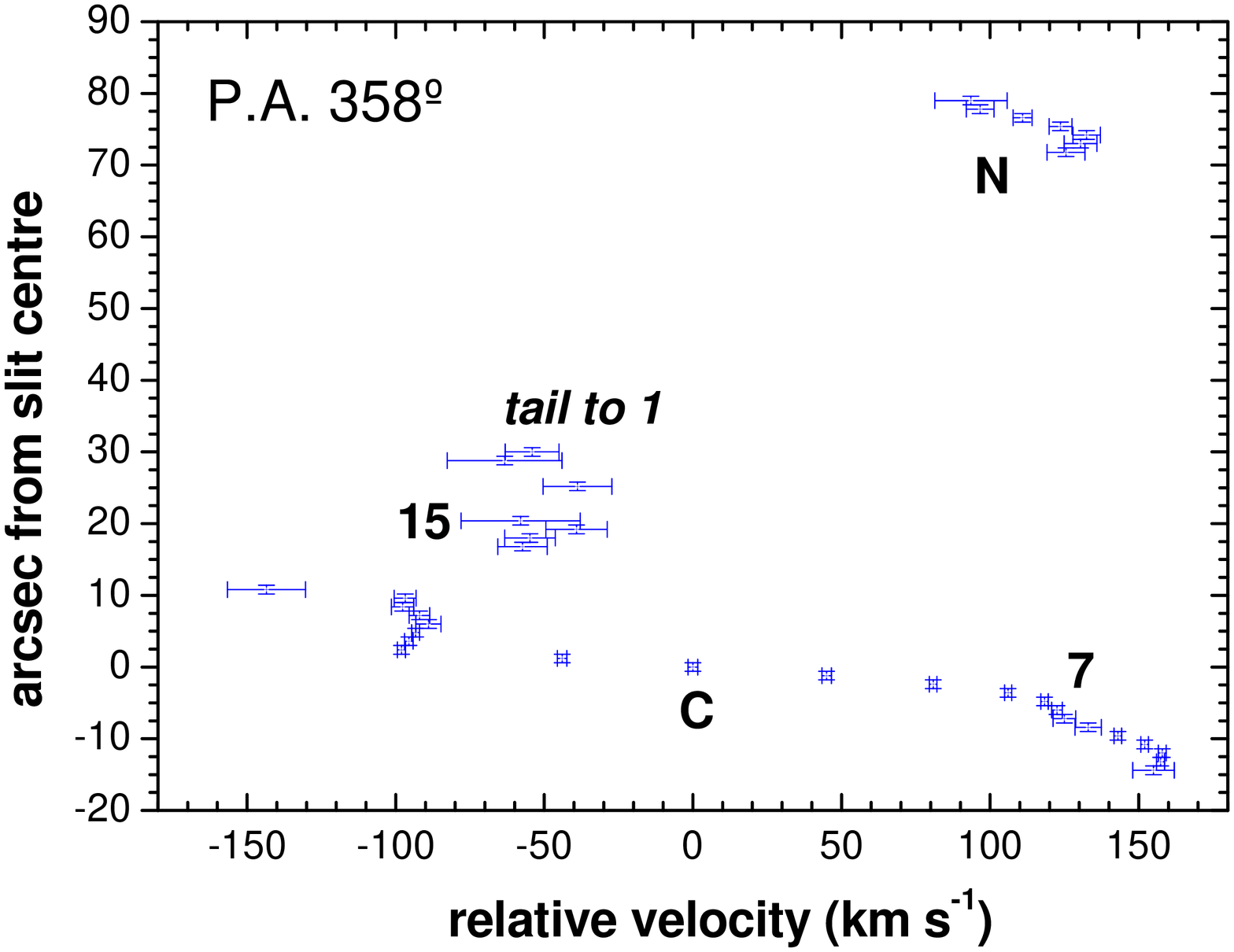}
\protect\caption[ ]{\small{Position-velocity diagram for the single slit position observed in Mkn 1087 with WHT, %%@
analysed in 1.2 arcsec bins. The horizontal bars represent the uncertainty of the Gaussian fitting for each point. The %%@
location and extension of the different galaxy members are also indicated. North is up.}}
\label{pv2}
\end{figure}

The kinematics of the ionized gas were studied via the spatially resolved analysis of bright emission line profiles %%@
(\Ha\ and \Hb\ for INT and WHT data, respectively) along each slit position. We have extracted zones of 4 pixels wide %%@
(1.6$\arcsec$ long) covering all the extension of the line-emission zones in the four INT slit positions shown in %%@
Figure~\ref{rendijas}, and zones of 6 pixels wide (1.2$\arcsec$ long) for the WHT slit position. The Starlink DIPSO %%@
software was used to perform a Gaussian fitting to the H$\alpha$ and H$\beta$ profiles. In figures~\ref{pv1} and %%@
\ref{pv2} we show the position-velocity diagrams for the four INT slit positions and the single WHT slit position, %%@
respectively. All the velocities are referred to the mean heliocentric velocity of the center of Mkn~1087 (8336 km %%@
s$^{-1}$). The position of the different objects is also indicated on figures~\ref{pv1} and \ref{pv2}. In %%@
Table~\ref{table3} we show the mean radial velocities obtained for each zone, that span a narrow interval of values %%@
with a maximum difference of about 250 km s$^{-1}$.

All the position-velocity diagrams show a more or less clear rotation pattern in the inner part of Mkn 1087. Knot \#7 %%@
participates of this rotation. However, the external zones have significant deviations from this behavior. Knots \#1 %%@
and \#3 are kinematically attached to the main body of the system despite their separation. We suggest that they are %%@
tidal dwarf galaxies (TDGs) formed from material stripped from Mkn 1087. Objects \#13, \#14 and \#15 seem to be %%@
kinematically attached to the rotation of Mkn 1087. There is a small slightly sinusoidal pattern in the gas velocity %%@
around \#13. This behavior, if real, could be produced by distortions associated with interaction effects \citep{S82, %%@
R90}. Note that knot \#1 has a velocity similar to \#13. The image of the system (see Figure~\ref{figR}) reveals an %%@
optical bridge between both objects, that furthermore have a very similar metallicity [12 + log(O/H) $\sim$ 8.50]. %%@
Figure~\ref{pv2} also shows some areas that can be interpreted as a tidal tail to knot \#1. All these evidences %%@
reinforce the idea that this structure (that has \#1 and \#13 at its extremes) could be actually a tidal tail.

The $HST$ image of the center of Mkn 1087 \citep{MGT98} shows a clear spiral pattern, implying a dynamically mature %%@
system. It also reveals that its west side is nearer to us. From the position-velocity diagrams of P.A. 357$^\circ$ %%@
and 358$^\circ$, it is clear that the north side of the spiral pattern is the approaching one. If these two diagrams %%@
are interpreted as a circular rotation, then we can estimate the Keplerian mass of Mkn 1087. Taking into account the %%@
half of the maximum velocity difference ($\sim$125 km s$^{-1}$) and the half of the spatial separation corresponding %%@
to these maxima ($\sim$12$\arcsec$ $\simeq$ 6500 pc), we obtain a mass of 2.37 $\times$ 10$^{10}$ \Mo\ for an %%@
inclination of 90$^\circ$, assuming circular orbits and Keplerian dynamics:
\begin{eqnarray}
\frac{M}{M_\odot} \sim 233 \times d~[\rm{pc}] \cdot \bigg(\frac{v~[\rm{km~s^{-1}}]}{\sin i}\bigg)^2
\end{eqnarray}
A similar result is found using the position-velocity diagram of P.A. 6$^\circ$. If we use P.A. 62$^\circ$ we estimate %%@
a velocity difference of around 100 km s$^{-1}$ in 8$\arcsec$ ($\simeq$ 4300 pc), obtaining a Keplerian mass of 1.0 %%@
$\times$ 10$^{10}$ \Mo, assuming an inclination of 90$^\circ$. If we consider that the elliptical shape of Mkn 1087 is %%@
only due to its inclination with respect the line of sight, we find that the proportion between the mayor axis (that %%@
correspond with P.A. 357$^\circ$) and the minor axis (between P.A. 62$^\circ$ and 120$^\circ$) of the assumed disk is %%@
5/3. This value corresponds with an inclination (defined as the angle between the plane of the sky and the plane of %%@
the galaxy) of 37$^\circ$ [\citet{GG81} derived an inclination angle of 38$^\circ$]. We now obtain a Keplerian mass of %%@
6.53 $\times$ 10$^{10}$ \Mo\ for P.A. 357$^\circ$ (mayor axis) and 4.63 $\times$ 10$^{10}$ \Mo\ for P.A. 62$^\circ$ %%@
(minor axis). For the last one, we have also corrected the projected distance, and it is the 70\% of the mass obtained %%@
from the position-velocity diagram of P.A. 357$^\circ$. The average value between both estimations gives a Keplerian %%@
mass of  $M_{\rm{kep}}$ = 5.6 $\times$ 10$^{10}$ \Mo\ for Mkn 1087. This value corresponds to a light to Keplerian %%@
mass ratio of $L_B/M_{\rm{kep}} \sim 2.19$. \citet{GG81}, using the Arecibo antenna, estimated a total hydrogen mass %%@
of  $M_{\rm{H\,I}}$ = 1.72 $\times$ 10$^{10}$ M$_\odot$ and a total dynamical mass of $M_{\rm{dyn}}$ = 1.78 $\times$ %%@
10$^{11}$ M$_\odot$ [from the velocity width of the \ion{H}{i} profile and applying the \citet{Br60} rotation law], %%@
that give a light to \ion{H}{i} mass ratio of $L_B/M_{\rm{H\,I}} \sim 7.12$ and a light to dynamical mass ratio of %%@
$L_B/M_{dyn} \sim 0.69$. Our $L_B/M_{\rm{kep}}$ ratio agrees with these estimates. Typical $L_B/M_{\rm{H\,I}}$ values %%@
for spiral galaxies are between 11.2 and 8.9 \citep{BGGB03}. Consequently, Mkn 1087 is a relatively gas-rich system.

\citet{K88} obtained position-velocity diagrams using various emission lines (H$\alpha$, H$\beta$, [\ion{N}{ii}], %%@
[\ion{O}{iii}]), instruments and detectors for P.A. 57$^\circ$ (this slit position crosses knot \#2) and for P.A. %%@
68$^\circ$. The last one is very similar to the position-velocity diagram we obtain for P.A. 62$^\circ$. The outer %%@
parts of the emission detected by \citet{K88} (that was chosen because it corresponds to one of the observed %%@
filaments) are redshifted by $\sim$200 km s$^{-1}$ with respect the inner part of Mkn 1087. He suggested that this %%@
feature indicates that this filament is infalling. However, as we will comment in \S 4.6, we think that the observed %%@
filaments in Mkn 1087 are tidal tails originated due to interaction with the nearby galaxy KPG 103a and the north %%@
companion object.

The companion object has a radial velocity similar to the main body of Mkn 1087 (+117 km s$^{-1}$), but it is %%@
kinematically detached of the rotation pattern of the main galaxy. 
In Figure~\ref{pv2} we can observe that the north companion object seems to have a sort of small rotation pattern, %%@
although its southern tip shows --if real-- a deviation that could be interpreted as a feature of the interaction %%@
between it and Mkn~1087. Assuming Keplerian rotation and an inclination of 90$^\circ$, we estimated a mass of 2.2 %%@
$\times$ 10$^8$ \Mo\ for this object, just two orders of magnitude lower than the value estimated for Mkn 1087. %%@
Typical values of the mass for dwarf galaxies are between 10$^7$ and 10$^8$ \Mo, so we can classify it as a dwarf %%@
object. These kinematic results, together with its low oxygen abundance ($\sim$8.24) reinforce the evidence that it is %%@
an external galaxy and not a TDG.   

We remark the fact that knot \#7 shows no important deviation of the main rotation pattern of Mkn 1087 in diagrams %%@
with P.A. 357$^\circ$ and 358$^\circ$. This fact does not support the \citet{K88} suggestion that \#7 is consequence %%@
of the merger of a dwarf galaxy. In the best case, this presumed merger should be in a very advanced stage. Our data %%@
is more consistent with the interpretation of knot \#7 as a giant \ion{H}{ii} region (or a complex of bright %%@
star-forming regions) located in one of the spiral arms of Mkn 1087.

\section{Discussion}

\subsection{Ages of the bursts and stellar populations}

\begin{figure*}[t]
\includegraphics[angle=270,width=0.5\linewidth]{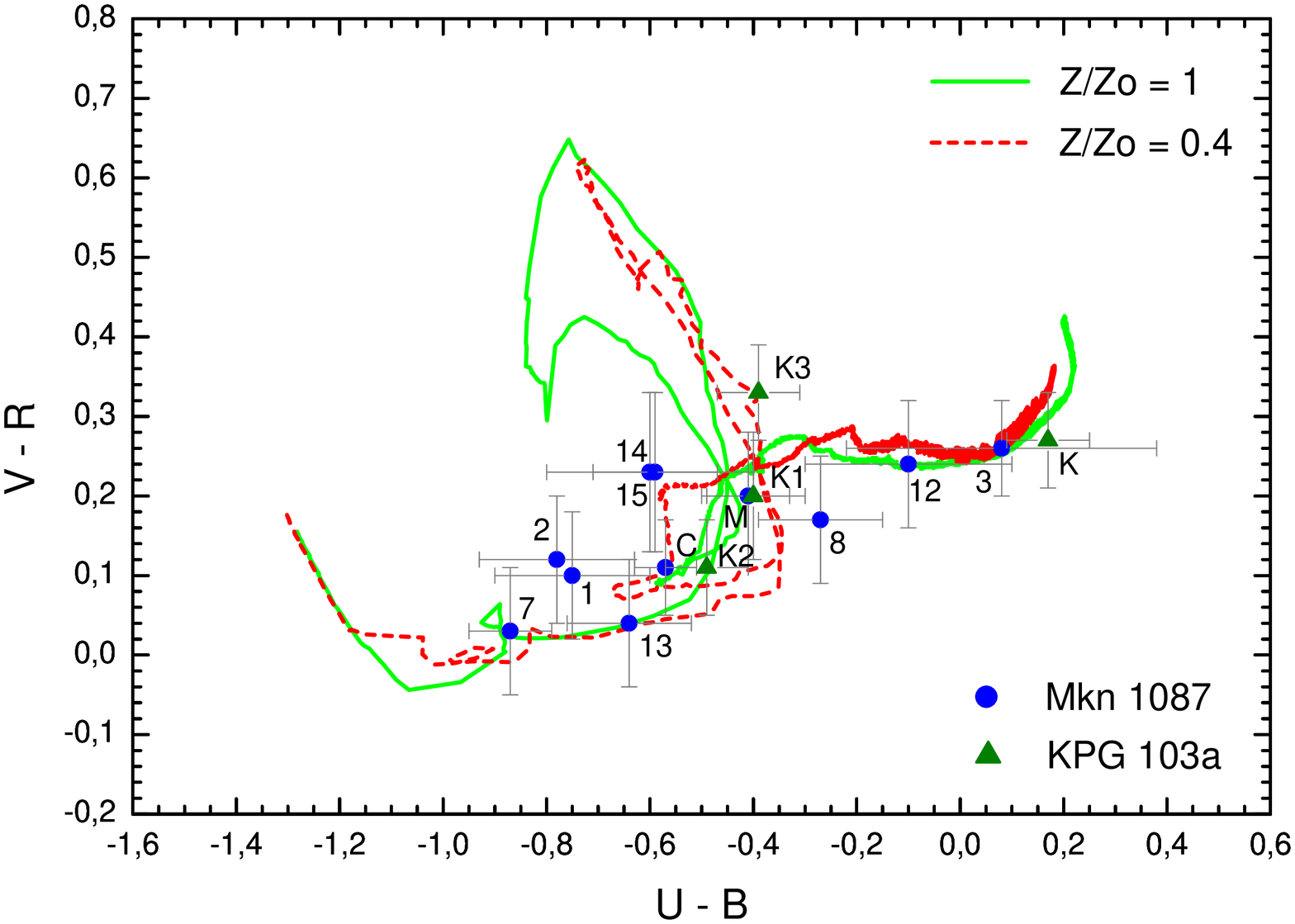}
\includegraphics[angle=270,width=0.5\linewidth]{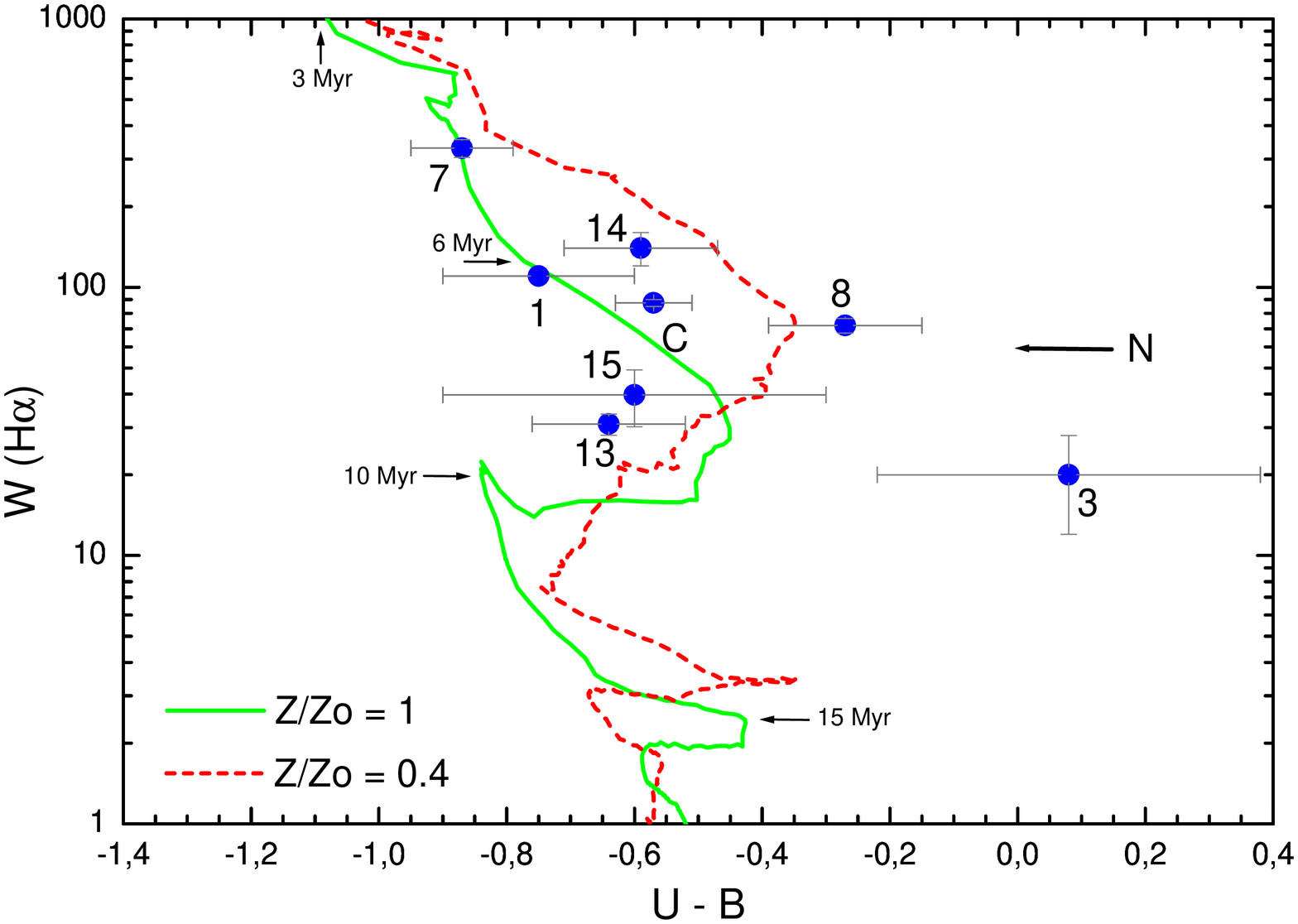}
\caption{\small{(a) $(V-R)$ versus $(U-B)$ values of the knots and STARBURST99 \citep{L99} predictions for an
instantaneous burst with a Salpeter IMF and $Z/Z_\odot$ = 1 and 0.4. The age of the bursts decrease from left to %%@
right. M represents the values for Mkn 1087 main body, whereas K represents the values for the galaxy KPG 103a. (b) %%@
$W$(\Ha) versus $(U-B)$ for knots in Mkn 1087 and
STARBURST99 \citep{L99} predictions. Some ages of the $Z/Z_\odot$ = 1 model are marked,
showing  the temporal evolution of the burst. The $W$(\Ha) of the north companion object is indicated with an %%@
horizontal arrow because there is not $U-B$ color available for it.}} 
\label{SB99}
\end{figure*}

ME00 used \citet{LH95} models to derive the ages of the bursts using their optical photometric data. We have improved %%@
their results combining our reddening-corrected broad-band optical and near-infrared photometric data with STARBURST99 %%@
\citep{L99} models (the improvement of Leitherer \& Heckman models). Spectral synthesis models with two different %%@
metallicities were chosen ($Z/Z_\odot$ = 1 and 0.4, the appropriate range of metallicities for the objects, see %%@
Table~\ref{table4s}), both assuming an instantaneous burst with a Salpeter IMF, a total mass of 10$^6$ $M_\odot$ and a %%@
100 $M_\odot$ upper stellar mass. The different color-color diagrams gave a similar interval of ages for all objects. %%@
In Figure~\ref{SB99}a the predictions of $(V-R)$ versus $(U-B)$ are compared with our observational values. Knots with %%@
redder colors (\#12, \#3 and the bright galaxy KPG 103a) correspond to the older objects in the system (knot \#11 %%@
should be included in this "red-group"). We have estimated an age of around 150 Myr for these three knots, values %%@
similar to those obtained by ME00.  

A detailed analysis of the procedure used to derive the age is described in our study of the Hickson Compact Group 31 %%@
\citep{LEM04}. We have also used the Wolf-Rayet starburst models by \citet{SV98} and \citet{SSL01} models to derive %%@
the age of the bursts using spectral features. A good agreement in all the age estimations has been obtained. We show %%@
the finally adopted value of the age of each spectroscopically studied burst in Table~\ref{table3}. We have found that %%@
all the starbursts are between 6 and 9 Myr old, except knot \#7 that seems to be younger (4.5 Myr). 

In Figure~\ref{SB99}b we plot our observational values of W(\Ha) (obtained from the spectroscopic data) versus the %%@
color (U$-$B) (derived from the photometric data), and the predictions of the STARBURST99 models. We find again a good %%@
agreement with the models except for knot \#3, an older object that hosts a little star-formation activity. 

As we commented before, the three bright knots (K1, K2 and K3) inside KPG 103a show blue colors [see their $(U-B)$ %%@
colors in Table~\ref{table2}]. Although we have no spectra for them, they seem to be starburst zones in the nucleus of %%@
the galaxy. We have used the same STARBURST99 \citep{L99} models to estimate the age of the star-formation in them. As %%@
it can be seen in Figure~\ref{SB99}a, the position of K1, K2 and K3 are consistent with the models, and we estimate an %%@
age between 6 and 7 Myr for them. However, if the color excess, $E(B-V)$, were higher than the one we have addopted %%@
(we only considered the Galactic contribution to the reddening for them, see \S3.1), these knots will show bluer %%@
colors and, consequently, they will be a slightly younger. 

\begin{figure}[t!]
\includegraphics[angle=270,width=1\linewidth]{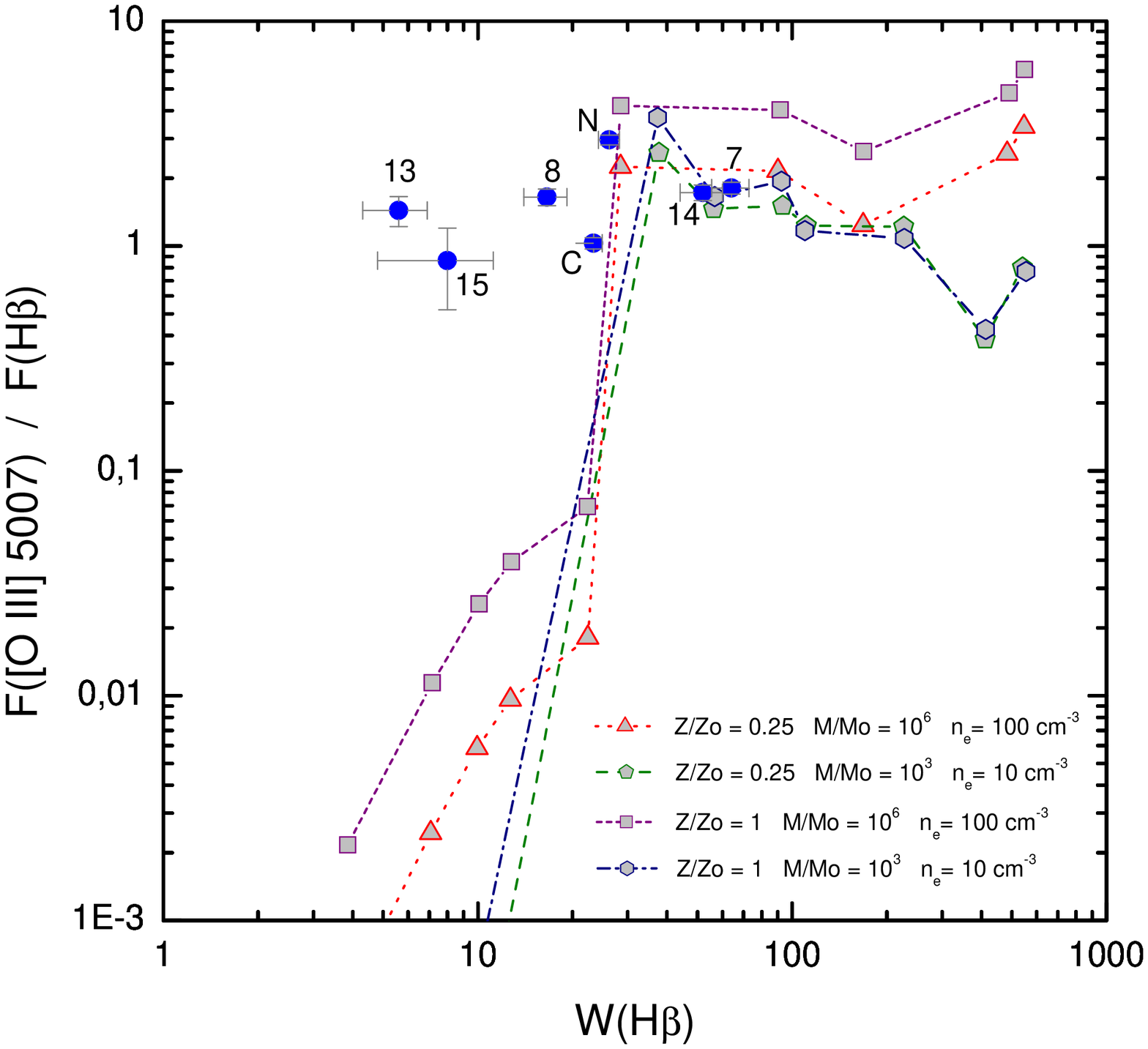}
\caption{\small{F([\ion{O}{iii}] $\lambda$5007) versus W (H$\beta$).
Models by \citet{SSL01}. Tracks correspond  to sequences of different
metallicities and electronic densities. Each symbol marks the position of the
models at 1 Myr interval, starting in the upper-right corner of the diagram
with an age of 1 Myr.}}
\label{sta}
\end{figure}

In Figure~\ref{sta} we plot our observational values of $W$(H$\beta$) and the [\ion{O}{iii}] $\lambda$5007/H$\beta$ %%@
emission line flux showed in Tables~\ref{table3} and \ref{table4a}, respectively, compared with the \citet{SSL01} %%@
models of \ion{H}{ii} regions ionized by an evolving starburst embedded in a gas cloud of the same metallicity. We %%@
indicate four models with two metallicities, $Z/Z_\odot$ = 0.25 and $Z/Z_\odot$ = 1, changing the total mass and %%@
electronic density of the ionized gas. All the objects show a good fit with the predictions, except knots \#13 and %%@
\#15. This fact can be due to a substantial population of old stars present in those objects, although it is difficult %%@
to detect them in the spectra. However, as we saw before, their colors and $W$(\Ha) indicate that they host a %%@
starburst that are around 8 Myr old.  

The spectrum of the center of Mrk 1087 (see Figure~\ref{fig6}a) shows clear \ion{H}{i} Balmer line absorptions, so we %%@
can estimate an effective age for the stellar population in the nucleus using the STARBURST99 \citep{L99} models. The %%@
best fit gives an age around 100 Myr, suggesting that star formation has been active in the nucleus of the galaxy ever %%@
since.

\begin{figure}[t]
\includegraphics[angle=270,width=1\linewidth]{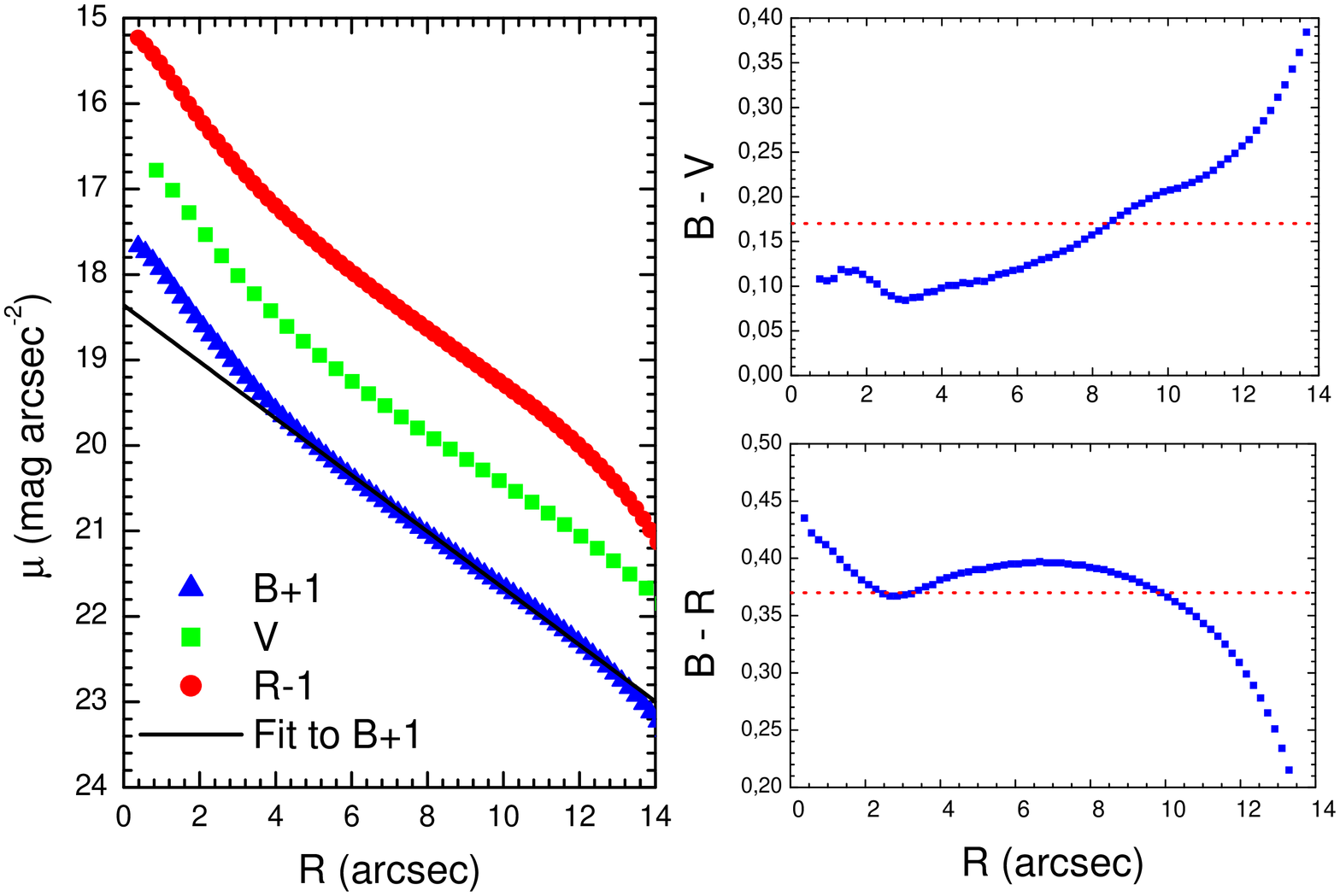}
\includegraphics[angle=270,width=1\linewidth]{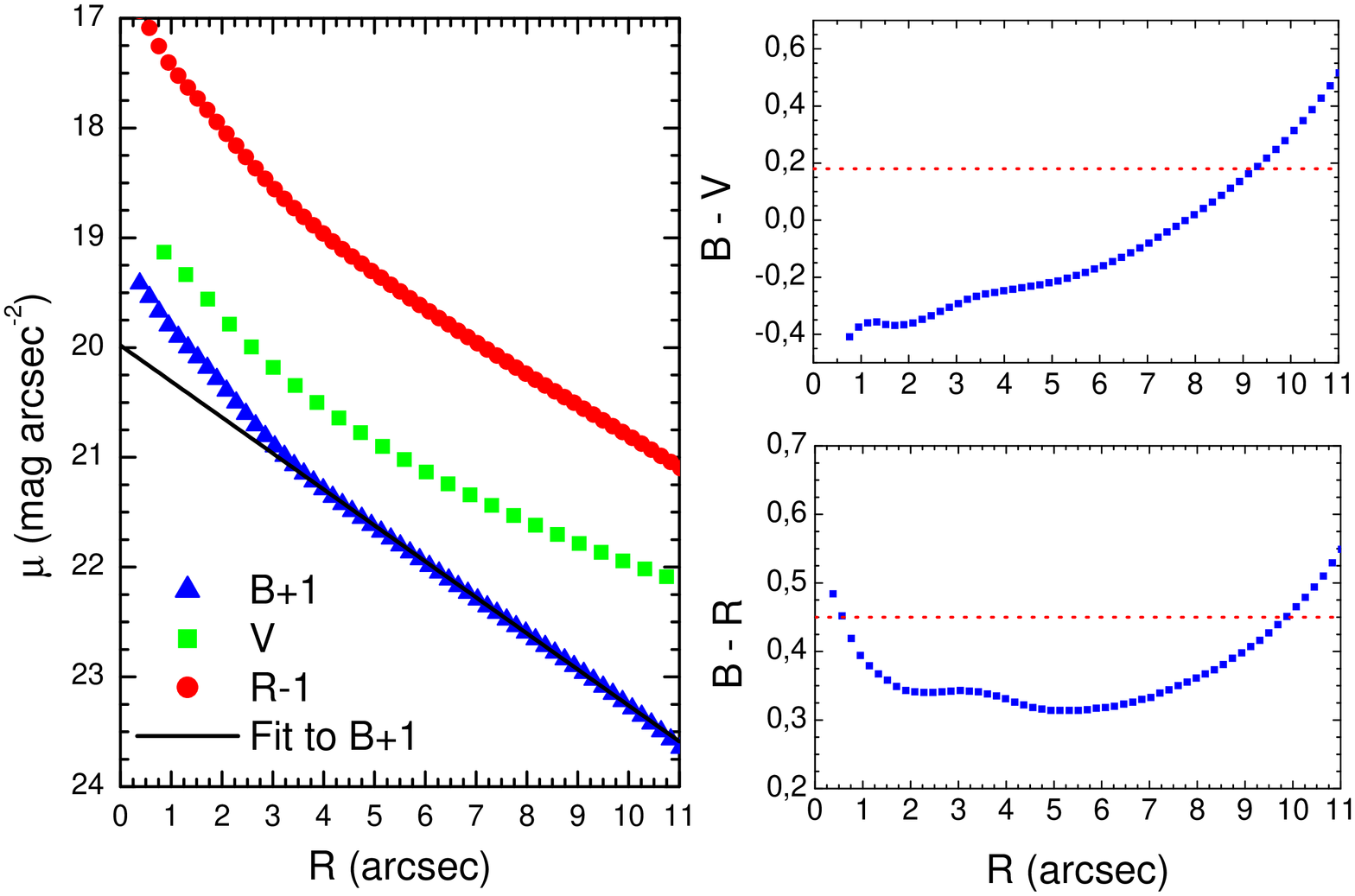}
\caption{\small{Surface brightness and $B$, $V$ and $R$ color profiles for Mkn 1087 (up) and KPG 103a (down). The line %%@
in the surface brightness diagrams is an exponential law fitting to the $B$ profile. The dotted horizontal line in the %%@
color profile diagrams indicates the average color derived for each system.}}
\label{p1}
\end{figure}

\begin{figure}[h]
\includegraphics[angle=270,width=1\linewidth]{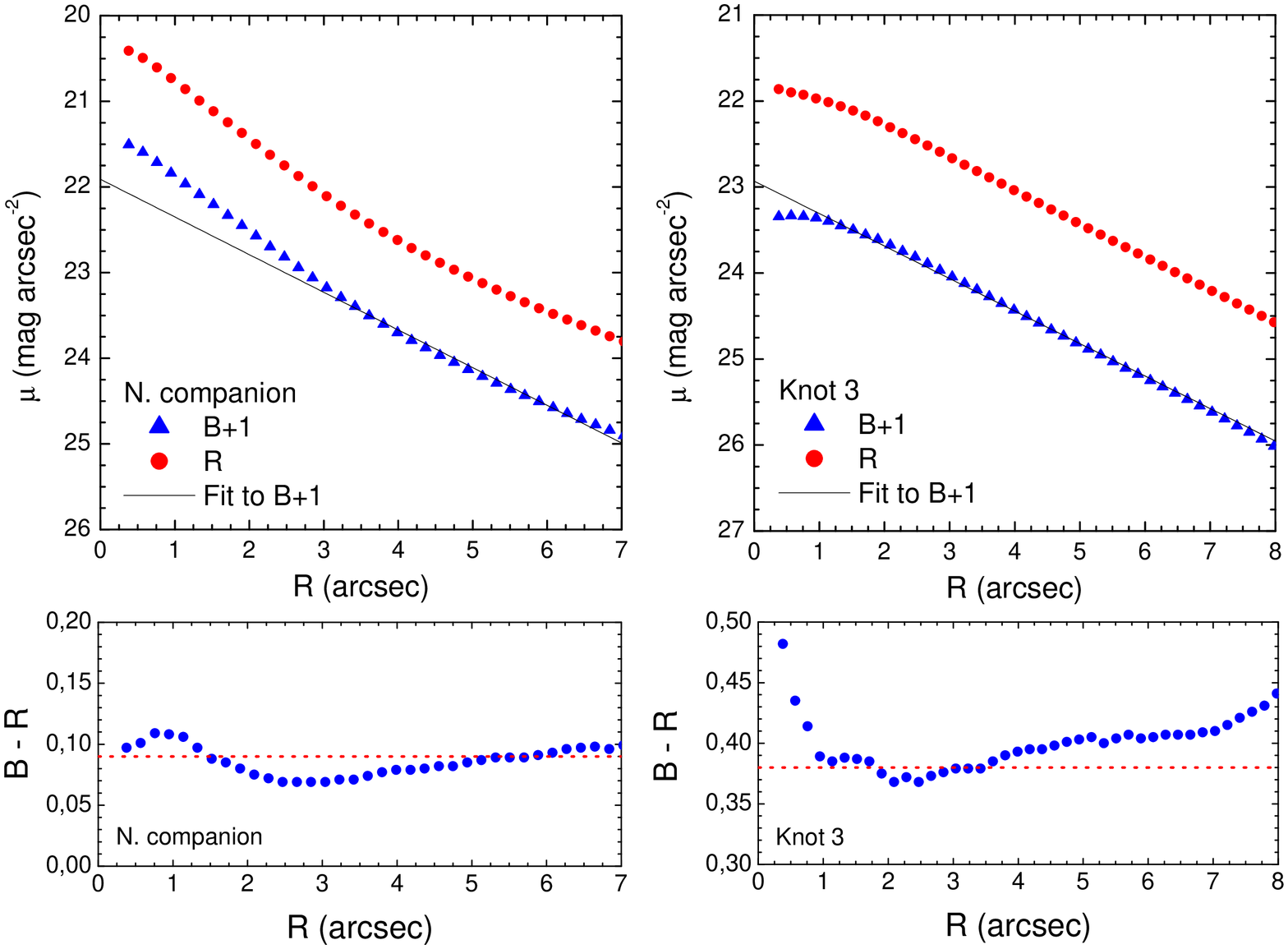}
\caption{\small{Surface brightness and color profiles for the north companion object and knot \#3. We follow the same %%@
notation than in Figure~\ref{p1}. Note that knot \#3 has not an underlying component.}}
\label{p2}
\end{figure}

In order to investigate the importance of older stellar populations, we have performed a simple analysis of the %%@
surface brightness profiles of the two main galaxies, as well as the north companion object and knot \#3. We have %%@
taken concentric surfaces at different radii from the center of each system, and
calculated the integrated flux inside each circle of area $A$ using the relation:
\begin{eqnarray}
\mu_X=m_X+2.5 \log A ,
\end{eqnarray}
to obtain the surface brightness, $\mu$ (in units of mag arcsec$^{-2}$), of each
circle $A$ (in units of arcsec$^2$) and $m_X$ is the magnitude in the filter $X$.  
In Figure~\ref{p1} we show the surface  brightness for $B$, $V$ and $R$ filters versus the
size of the aperture (in arcsec) for Mkn 1087 and KPG 103a, whereas the surface brightness for the north companion and %%@
object \#3 are shown in Figure~\ref{p2}. In these two figures we also show the radial color profiles $(B-R)$ and %%@
$(B-V)$ (this last only in Figure~\ref{p1}) derived by a direct subtraction of light profiles. The dotted horizontal %%@
line indicates the representative average color derived for each system (see Table~\ref{table2}).

\begin{table*}[t!]\centering

  \caption{Structural parameters of Mkn 1087 and the brightest galaxies and objects associated with it. }
  \label{table7}  
  \small
  \begin{tabular}{cccccccc}
   \noalign{\smallskip}
    \hline\hline
	\noalign{\smallskip}
   Member &   U.C.$^a$& $\mu_{B,0}$& $\alpha_B$ (kpc) & $\mu_{V,0}$& $\alpha_V$ (kpc) & $\mu_{R,0}$& $\alpha_R$ (kpc) %%@
\\
    \hline
	\noalign{\smallskip}
 	Mkn 1087 & yes&  17.36     &  1.63	& 17.34  &  1.72   & 16.85  &  1.55  \\
	KPG 103a & yes&  18.98     &  1.64  & 19.78  &  2.44   & 18.77  &  1.76 \\
	N. comp. & yes&  20.91     &  1.22	&\nodata & \nodata & 20.89  &  1.22   \\
	\#3      &  no&  21.93     &  1.42	&\nodata & \nodata & 21.52  &  1.41   \\
   \hline\hline
  \end{tabular}
  \begin{flushleft}
  $^a$ Underlying component.\\
  \end{flushleft}
   
 \end{table*}
 
We can observe that the surface brightness profile can be separated in two structures in all objects except \#3, %%@
especially in the case of the $B$-band, indicating that they have a low surface brightness component underlying the %%@
starburst. In the case of the two brightest galaxies, we can attribute this underlying component to the galactic disk. %%@
We have performed an exponential law fitting to the profiles, following the
expression:
\begin{eqnarray}
I=I_o \exp(-\alpha \, r),
\end{eqnarray}
that describes a typical disk structure: $I_0$ is the central intensity and
$\alpha$ is the scale length. The fitting structural parameters are indicated
in Table~\ref{table7} and the fit is plotted over each $B$ profile with a straight line.
It is not a surprise to find two structural components in Mkn 1087 and KPG 103a: Figure~\ref{fig3} indicates the %%@
presence of a galactic disk with a possible spiral structure in both galaxies. Furthermore, it is interesting to %%@
remark the noticeable variation in the $(B-V)$ color along KPG 103a: it seems that the star-formation activity is very %%@
concentrated at its center, probably in the three inner knots (K1, K2 and K3) analyzed in \S3.1. 
Note that they show considerably bluer colors, $(U-B)\sim-0.4$, than the one derived for the complete galaxy, $(U-B)$ %%@
= 0.17. We also find two structures in the north companion object (Figure~\ref{p2}), indicating that an important old %%@
population is present and further supporting the idea that this object actually is an external galaxy nad not a tidal %%@
dwarf. However, we do not find two components in knot \#3. Is the contribution of the starburst dominating the %%@
underlying population? This problem of the "missing populations" in starbursts can be solved using high spatial %%@
resolution photometry in NIR \citep{N03}. However, we do not think that it could be applied in \#3: its low %%@
star-formation activity, colors and age suggest that most of the structure observed corresponds to an old stellar %%@
population. The weak contribution of the starburst is perhaps, in this particular case, diluted under the dominant old %%@
component.       

\subsection{Star formation rates \\ and the classification of Mkn 1087}

Mkn 1087 and KPG 103a were detected jointly by IRAS (\emph{InfraRed Astronomical Satellite}) at 25, 60 and 100 $\mu$m %%@
(source F04470+0314), with fluxes of $f_{25}$ = 0.50 Jy, $f_{60}$ = 3.48 Jy and $f_{100}$ = 4.77 Jy, respectively %%@
(IRAS Point Source Catalog, 1986). Using the \citet{K98} calibration, we find that the star formation rate (SFR) %%@
derived for the system from infrared fluxes is SFR$_{\rm{IR}}$=11.5 \Mo\ yr$^{-1}$. It is a typical value for star %%@
formation galaxies. We can also estimate the dust temperature, $T_D$, and the warm dust mass, $M_D$, using the %%@
expressions given by \citet{BGGB03} and the $f_{60}$ and $f_{100}$ fluxes. We obtain $T_D = 43.2$ K and $M_D = 7.62 %%@
\times 10^6$ \Mo. Considering the luminosity of both galaxies, it corresponds to a mass to light ratio of $\log M_D / %%@
L_B = -4.36$, a little lower than the expected values for spirals [between $-$3.74 and $-$4.07, \citep{BGGB03}].  

Radio emission from galaxies is an extinction-free tracer of star formation. \citet{CCB02} derived $\log L_{1.4\, %%@
\rm{GHz}}$ = 22.33 for Mkn 1087 using the NRAO VLA Sky Survey (NVSS). We can estimate the SFR from the 1.4 GHz %%@
luminosity using this data and the calibration given by \citet{CCB02}. We obtain a value of SFR$_{1.4\, \rm{GHz}}(M > %%@
5M_{\odot}) \sim$ 4.65 \Mo\ yr$^{-1}$, that could be considered as the rate of massive star formation during the past %%@
$10^8$ yr. Assuming a Salpeter initial mass function between 0.1 and 100 \Mo, we can extrapolate it to include lower %%@
masses, finding SFR$_{1.4\, \rm{GHz}}(M > 0.1M_{\odot}) \sim$ 25.8 \Mo\ yr$^{-1}$, a relative high value compared with %%@
the SFR derived from FIR fluxes. This is a reasonable result because SFR$_{1.4\, \rm{GHz}}$ measures the SFR of the %%@
most recent bursts of the star formation, while SFR$_{\rm{IR}}$ indicates the average SFR of a stellar population with %%@
larger average life times.

The SFR can be also derived using the H$\alpha$ emission, applying the calibration given by \citet{K98}. Although we %%@
do not have H$\alpha$ imagery of the system, we can make an estimation of the H$\alpha$ from our spectra considering %%@
the slit size with respect to the total area of the bursts. We have estimated that the H$\alpha$ flux (reddening %%@
corrected) we have measured in C is $\sim$ 10 \% of the total of the galaxy. This value corresponds to SFR$_{H\alpha} %%@
\sim$ 4.5 \Mo\ yr$^{-1}$ for Mkn 1087. As we expect, this approximated value is lower than the ones derived from FIR %%@
and 1.4 GHz fluxes because it does not take into account the dust-obscured areas. The probable range for the real SFR %%@
in Mkn 1087 is between 4.5 (from H$\alpha$ flux) and 11.5 (from IR fluxes) \Mo\ yr$^{-1}$.

\begin{figure}[t!]
\includegraphics[angle=270,width=1\linewidth]{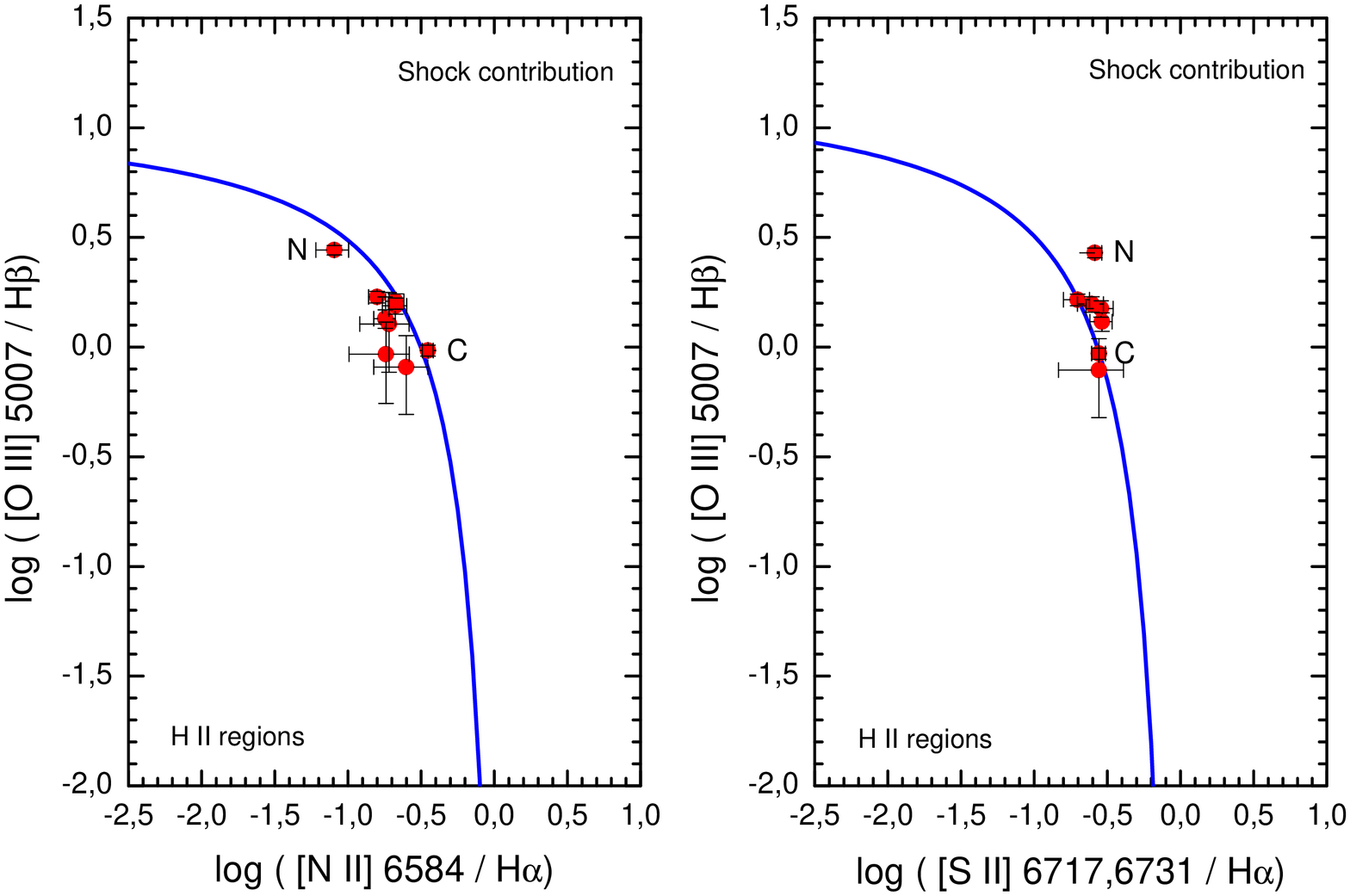}
\caption{\small{Emission line ratios of knots in Mkn 1087. The solid lines give the limit for ionization by a zero age %%@
starburst, following \citet{Do00}.}}
\label{dopita}
\end{figure}

Some authors, as \citet{CCB02}, classified Mkn 1087 as Active Galactic Nucleus (AGN) but our data do not support it. %%@
The spectrum of the central zone of Mkn 1087 (see Figure~\ref{fig6}) is similar to the ones of typical starburst %%@
galaxies. The FWHM of the \Hb\ and [\ion{O}{iii}] $\lambda$5007 emission lines are 223 km s$^{-1}$ and 228 km s$^{-1}$ %%@
(corrected for instrumental broadening), respectively. The emission line widths in nuclear \ion{H}{ii} regions and %%@
starburst galaxies have values up to $\sim$300 km s$^{-1}$, whereas the typical FWHM median value range for AGNs is %%@
between 350 and 550 km s$^{-1}$ \citep{VG97}. Furthermore, we can make use of the \citet{Do00} diagnostic diagrams %%@
based on emission line ratios to classify the excitation mechanism of the emission-line systems. In %%@
Figure~\ref{dopita} these relations are plotted together with our observational data, being evident that the knots %%@
studied here are consistent with the loci of typical \ion{H}{ii} regions and not with AGNs. AGNs are clustered around %%@
$\log($[\ion{O}{iii}]/\Hb$)\sim$ 1, $\log($[\ion{N}{ii}]/\Ha$)\sim$ 0.1 and $\log($[\ion{S}{ii}]/\Ha$)\sim$ $-$0.2 %%@
\citep{VO87}.   

However, Mkn 1087 could be classified as a Luminous Compact Blue Galaxy (LCBG). LCBGs are $\sim L^{\star}$ ($L^{\star} %%@
= 1.0 \times 10^{10} L_{\odot}$), blue, high surface brightness, high metallicity, vigorous starbursting galaxies with %%@
an underlying older stellar population \citep{GJK98}. Specifically, a LCBG has $(B-V) < 0.6$, a surface brightness %%@
$\mu_B <$ 21 mag arcsec$^{-2}$ and $M_B\,<\,-$18.5. Mkn 1087 satisfies all these properties (for a radius of 5 arcsec, %%@
it has $(B-V)$ = 0.11, a $\mu_B\,\sim$ 21 mag arcsec$^{-2}$ and $M_B$ = $-$22.3). LCBGs are quite common at %%@
intermediate redshifts, but by $z\, \sim\, 0$ their number density has decreased by a factor of ten and seem to have, %%@
on average, an order-of-magnitude smaller masses that typical galaxies today with similar luminosities \citep{GOK03}. %%@
The observed properties in LCBGs at low $z$ are similar to the high $z$ Lyman-break galaxies \citep{EP03}. %%@
Consequently, LCBGs are especially interesting for studies of galaxy evolution and formation because they could be the %%@
equivalent of the high $z$ Lyman-break galaxies in the local universe and they have evolved more than other galaxy %%@
class. 

\subsection{WR population}

We have not detected neither the blue Wolf-Rayet bump nor the nebular \ion{He}{ii} $\lambda$4686 emission line in the %%@
spectrum of the central zone of Mkn 1087. However, we have used the data obtained by \citet{Va97} to derive the WR %%@
population in the galaxy, assuming that all the flux came from the \ion{He}{ii} $\lambda$4686 emission line. These %%@
authors estimated that $F$(\Hb) = 7.81$\times10^{-14}$ erg s$^{-1}$ cm$^{-2}$ and $F$(\ion{He}{ii} $\lambda$4686) = %%@
8.36$\times10^{-15}$ erg s$^{-1}$ cm$^{-2}$. If we consider that all the contribution of the \ion{He}{ii} %%@
$\lambda$4686 emission line comes from WNL we find around 7200 WNL stars assuming that $L$(WNL 4686) = %%@
1.7$\times10^{36}$ erg s$^{-1}$ for a WNL star \citep{VC92}. This value is somehow intriguing, because \citet{VC92} %%@
reported an upper limit of 1600 to the total WNL presented in Mkn 1087, a factor 4.5 smaller than our estimate. Are %%@
\emph{really} there Wolf-Rayet stars? The slit used by \citet{Va97} was 2.4$\arcsec$ $\times$ 4.8$\arcsec$, whereas %%@
the one used by \citet{VC92} was 1.5$\arcsec$ wide, but they do not indicate its long. We used a 1$\arcsec$ $\times$ %%@
5.4$\arcsec$ aperture in our WHT observations. Perhaps, the WR stars are concentrated in a very small area and %%@
\citet{Va97} were the only ones observing at the correct place.

To derive the WR/(WR+O) ratio, we have to correct the contribution of the WR stars to the ionizing flux to obtain the %%@
total number of O stars, assuming that $L$(\Hb) = 4.76$\times10^{36}$ erg s$^{-1}$ for a O7V star \citep{VC92} and %%@
$\eta\equiv$ O7V/O = 0.25 \citep{SV98}. We derived around 47500 O stars in Mkn 1087, that implies a WR/(WR+O) ratio of %%@
0.13. We have also used the calibration of the WR/(WR+O) ratio using the flux of the blue WR bump given by %%@
\citet{SV98} in their study about evolutionary synthesis models for O and WR populations in young starburst (their %%@
equation 17). Assuming that all the blue WR bump flux comes from the \ion{He}{ii} $\lambda$4686 emission line and %%@
using the \ion{He}{ii} $\lambda$4686 flux from \citet{Va97}, we estimate a WR/(WR+O) ratio of 0.12, a value in %%@
excellent agreement with our previous determination. However, these values represent an upper limit to the real one %%@
because \citet{Va97} did not specify the stellar or nebular origin of the \ion{He}{ii} $\lambda$4686 emission line.  

The derived age (see \S4.1 and Table~\ref{table3}) for the youngest bursts of star formation in Mkn 1087: knots C (the %%@
central zone) and \#7, between 4 and 6 Myr, suggest that WR stars should be present in so young massive starbursts. %%@
Aperture effects seem to be playing an important role in the no-detection of WR features. 

\subsection{The chemical composition of the objects.}

\begin{figure}[t!]
\includegraphics[angle=270,width=1\linewidth]{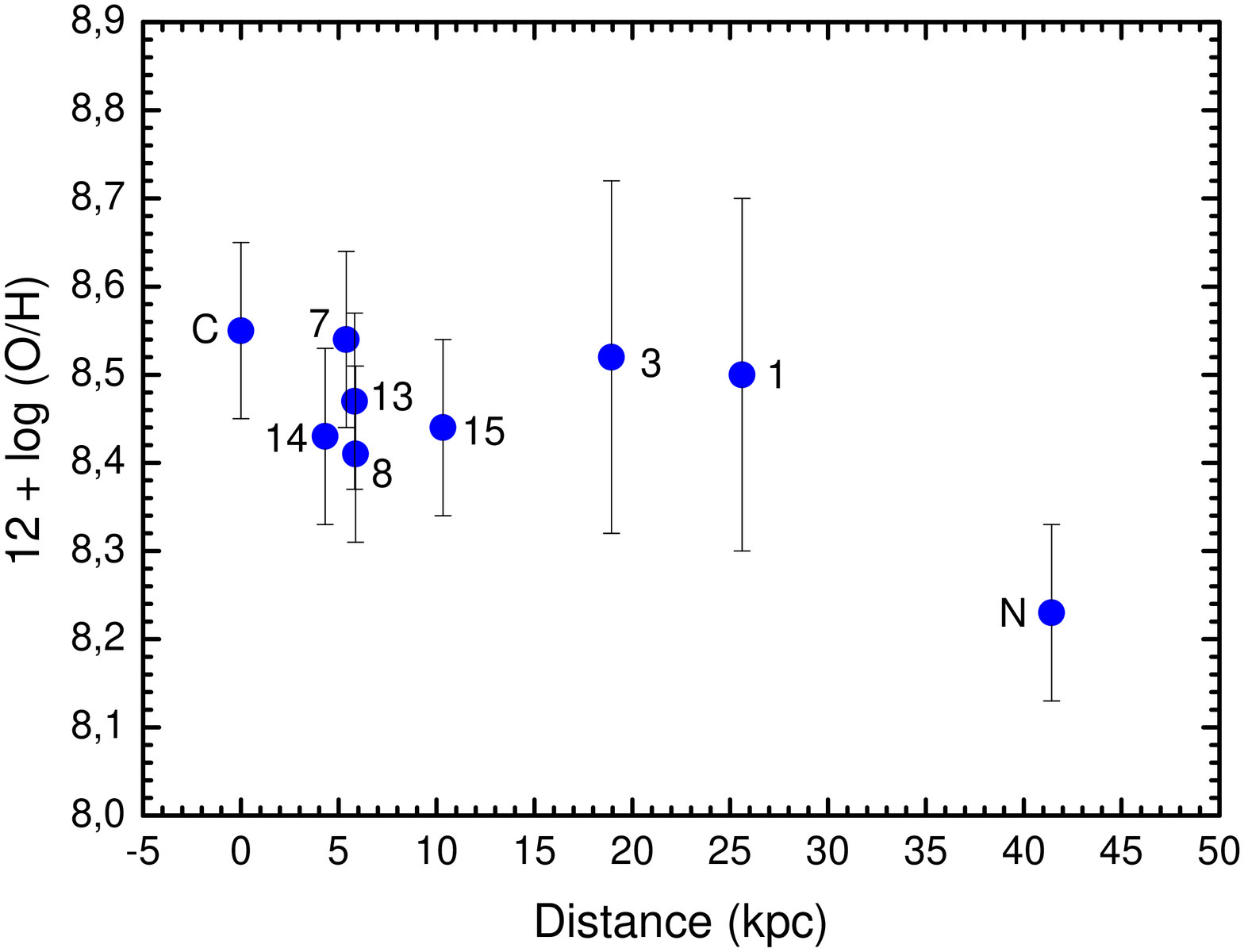}
\caption{\small{Relation between metallicity and distance from the nucleus of Mkn 1087 for all the observed knots. We %%@
have considered the oxygen abundances derived using the calibration by \citet{P01}.}}
\label{radial}
\end{figure}

We have analyzed the radial dependence of the derived metallicity of the knots in Mkn 1087, which is shown in %%@
Figure~\ref{radial}. We have used the oxygen abundance obtained from the empirical calibration of \citet{P01} (see %%@
Table~\ref{table4s}), although this method gives a high uncertain oxygen abundance for knots \#1 and \#3 (we have %%@
assumed an error of $\pm$0.20 dex for both, and $\pm$0.10 dex for the rest of the objects, see \S 3.2.2.) 

We can distinguish two zones in Figure~\ref{radial}: the first one is the galaxy disk itself, that extends to around %%@
10 kpc from the center of Mkn 1087. We find 5 different objects in this area (the inner knots), two of them (C and %%@
\#7) observed both with INT and WHT (we use the average value of both independent determinations). The object \#15 %%@
lies just at the border of the disk. The second zone is the external area, where the north companion galaxy and knots %%@
\#1 and \#3 are located.

We have explained above that knot \#7 seems to be an intense star-formation zone off-center of Mkn 1087. It is %%@
resolved into numerous individual knots in the $HST$ image presented by \citet{MGT98}. Figure~\ref{radial} suggests %%@
that the oxygen abundances of the center of the galaxy and \#7 are very similar, despite the distance of 6 kpc between %%@
them. \citet{K88} suggested that \#7 is the consequence of a merger, but the position-velocity diagrams for P.A. %%@
357$^\circ$ and 358$^\circ$ do not show any dynamical evidence of this. Furthermore, the individual knots that form %%@
\#7 in the $HST$ image do not show a clear central concentration, supporting the idea that it is an %%@
interaction-triggered star-forming region and not the core of a stripped companion \citep{WHM96}. The oxygen abundance %%@
of the different knots along the disk of Mkn 1087 is rather similar, but the central zone shows a slightly higher %%@
value [12+log(O/H) = 8.55, whereas the external zones show a values around 8.45]. On the other hand, the N$^+$/O$^+$ %%@
ratio is about 0.2$-$0.3 dex higher in the nucleus. These results suggest the possible presence of a rather weak %%@
abundance gradient along the disk of Mkn 1087 (about $-$0.013 dex kpc$^{-1}$), although nothing definitive can be said %%@
taking into account the uncertainties. Typical values of oxygen gradients in spiral galaxies are between $-$0.009 dex %%@
kpc$^{-1}$ and $-$0.231 dex kpc$^{-1}$, with an average gradient of $-$0.06 dex kpc$^{-1}$ \citep{ZKH94}.

The external knots \#1 and \#3 are located at 26 and 19 kpc (respectively) from the center of Mkn 1087 and seem to %%@
show oxygen abundances similar to the ones obtained in the disk of the galaxy [12+log(O/H) $\sim$ 8.50]. Taken at face %%@
value, this gives further evidence that knots \#1 and \#3 are tidal dwarf galaxies. However, the high uncertainties in
their metallicities preclude any definitive conclusion from simple abundance considerations.

From Figure~\ref{radial}, it is clear that the north companion object has a significant lower O/H ratio than the disk %%@
of Mkn 1087. The chemical difference is also evident in the N$^{+}$/O$^+$ ratio. N$^{+}$/O$^+$ is about $-$1.4 in the %%@
companion and between $-$0.8 and $-$1.1 in the rest of the knots. The value of $-$1.4 is of the order of that typical %%@
for dwarf galaxies \citep{G97}. This is a further evidence that the north companion is an external pre-existing %%@
galaxy.

\subsection{The luminosity-metallicity relation}

\begin{figure}[t]
\includegraphics[angle=270,width=1\linewidth]{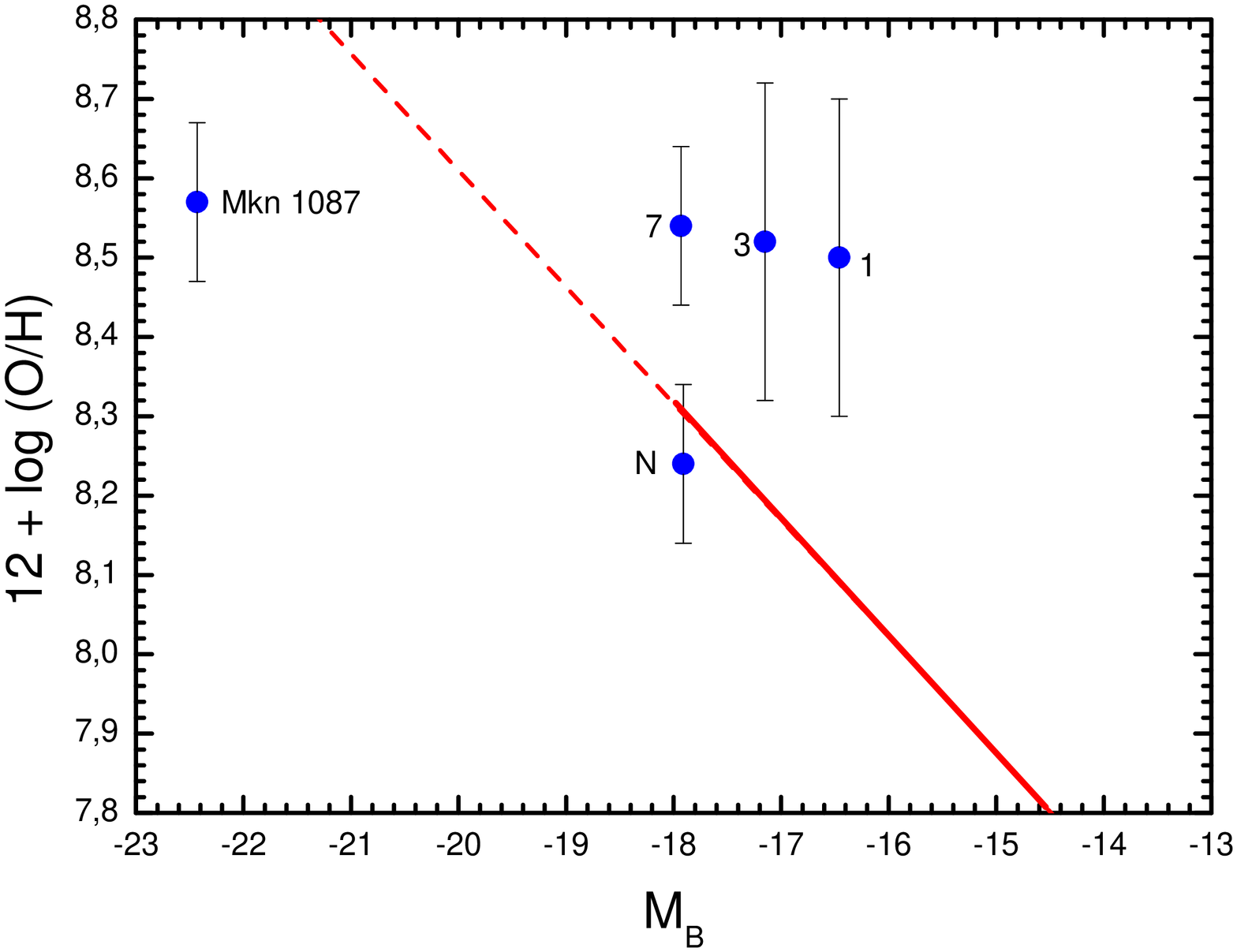}
\caption{\small{Luminosity-metallicity relation for
the observed detached bursts in Mkn 1087 and knot \#7. The solid line is the relation for dwarf irregulars
found by \citet{RM95}, while the dashed line is an extrapolation of it. We have considered the oxygen abundances %%@
derived using the calibration by \citet{P01}.}}
\label{richer}
\end{figure}

\citet{RM95} obtained the following relation between absolute $B$ magnitude and oxygen abundance for dwarf irregular %%@
galaxies (for $M_B$ $\geq$ $-$18, the most common definition for a dwarf galaxy):
\begin{eqnarray}
12+\log {\frac{\rm{O}}{\rm{H}}} = 5.67 - 0.167 \times M_B
\end{eqnarray}
that we have used for the objects in Mkn 1087. In Figure~\ref{richer} we show this relation (with an extrapolation to %%@
higher luminosities). We have assumed that Mkn 1087 has the metallicity of its center, but obviously its position is %%@
under the \citet{RM95} relation because it is not a dwarf galaxy. 

In Figure~\ref{richer} we can observe that the position of the north companion is consistent with the %%@
metallicity-luminosity relation for dwarf galaxies (in fact, it is the only object that is consistent with it). %%@
Therefore, the north companion should be interpreted as an independent nearby dwarf galaxy. The presence of the tidal %%@
dwarf \#1 and its associated bridge, which is almost aligned with the companion galaxy, suggests that the north %%@
companion is in interaction with Mkn 1087.

\citet{DM98} found that dwarf galaxies with metallicities higher than the ones expected from the %%@
metallicity-luminosity relation (i.e, dwarf galaxies located at the upper-right zone in Figure~\ref{richer}) are TDGs. 
Knots \#1 and \#3 are located in this zone of the diagram and away from the \citet{RM95} relation for dwarf irregular %%@
galaxies. This behavior is consistent with their possible TDG nature, in the sense that they have a much larger %%@
metallicity than it is expected for normal galaxies of the same luminosity. This evidence reinforces our hypothesis %%@
that both objects are tidal dwarf galaxies formed from material stripped by Mkn 1087, perhaps due to the interaction %%@
with the dwarf north companion object (in the case of \#1 seems likely) and maybe also with the neighbour galaxy KPG %%@
103a. 

Finally, the position of knots \#7 in the metallicity-luminosity relation does not support the claim by \citet{K88} %%@
that this object is a merged external galaxy. If this is true, the position of the object should be more consistent %%@
with the relation.

\subsection{Mkn 1087: A probable group in interaction.}

Sometimes, interaction features are found in group of galaxies, especially in Hickson Compact Groups [see, for %%@
example, \citet{IPV01}; HCG~92: the \emph{Stephan's Quintet} \citep{Y97, SRV01}; HCG~54 \citep{VM02}; NGC 6845 galaxy %%@
group \citep{SKJ03}; HCG~31 \citep{LEM04}]. We consider that Mkn 1087, KPG 103a and the north companion object are, in %%@
fact, a group of galaxies. We might also include the tidal object \#3 in this group.

We have studied the selection criteria given by \citet{H82} and \citet{H92} to define a compact group of galaxies in %%@
the particular case of Mkn 1087. Our group satisfied the criteria of the number of members (at least, three galaxies %%@
that have similar radial velocities), isolation (there is not other galaxies with $\Delta mag<$3 in a radii lower than %%@
3 times the assumed size of the group) and compactness criterion (the mean surface brightness of the group, $\mu_G$, %%@
defined as the distribution of the flux of all the members over the size of the group, is around 23.5 and, therefore, %%@
lower than 26). However, it fails in the relative magnitude criterion because the north companion galaxy differs in %%@
more than 3 magnitudes from the absolute magnitude of the brightest galaxies. Consequently, Mkn 1087 and its %%@
surroundings can not be strictly classified as a compact group of galaxies attending to Hickson's criteria, which also %%@
are defined by the limitations of Hickson's survey. What it is clear to us is that they actually form a group of %%@
galaxies with, at least, three members (four, if we also consider object \#3). 

The knowledge of the dynamics and galaxy orientations in the group offers some clues to understand what kinds of %%@
interactions can lead to the observed structures. Mkn 1087 is a spiral galaxy with an inclination of $\sim 37^\circ$ %%@
with respect to the visual. Its west side is nearer to us and the north side is the approaching one (see \S3.2.3). All %%@
the tails seem to be nearly radial to the spiral disk, but the streamers pointing toward objects \#11 and \#12 might %%@
be roughly polar, while the tail to knot \#1 could be close to the disk plane. It is intriguing that intense burst \#7 %%@
is also aligned in this direction. 

Knot \#15, located at the beginning of the long tail to object \#1, shows blue colors, $(U-B)\sim -0.6$, indicating %%@
star formation activity. The images (see Figures~\ref{figR} and \ref{rendijas}) reveal small knots along this tail, %%@
that decrease in number in the direction to \#1. They seem to be tiny star-forming regions too, because very weak \Ha\ %%@
emission is detected in this zone of the WHT spectrum. Knot \#1 appears to be an irregular object composed by diffuse %%@
little knots spanning about 4$\arcsec$. We do not think it is a disrupted pre-existing object instead of a tidally %%@
stripped remnant because its metallicity is very similar to the spiral disk of Mkn 1087. Furthermore, if knot \#1 were %%@
a disrupted low-mass object, its gravitational potential would be strong enough to modify considerably the spiral %%@
pattern of Mkn 1087 \citep{WHM96}. In "minor" mergers, the main galaxy develops a strong two-armed spiral pattern, %%@
whereas the nucleus of the low mass external satellite will finally be a small bulgelike entity. Examples of minor %%@
mergers are Arp 31 \citep{vM88}, NGC 7252 [the \emph{Atoms-for-Peace} galaxy \citep{S82,HGGS94}] or NGC 1097 %%@
\citep{HW03}. All these systems show that the inner part of the main galaxy is very disturbed with ripples, loops and %%@
tails. None of these features are observed in our case: the $HST$ image of Mkn 1087 reveals a clearly and undisturbed %%@
spiral pattern, something that is also observed in the position-velocity diagrams (see Figures~\ref{pv1} and %%@
\ref{pv2}), and knot \#1 is not a bulgelike object. Moreover, \citet{MH94} proposed that these kind of mergers can %%@
lead to a significant inflow of disk gas and fuel a strong central starburst in the main galaxy, being predominantly %%@
concentrated in the inner kiloparsec of the disk rather than throughout the disk or bulge. This is not observed in %%@
Mkn1087; the star-formation activity is found all over the disk.

The new data presented through our study give some clues to understand the evolution of the group. The number and %%@
intensity of tails and bridges that are observed indicate that interaction processes have been important in the recent %%@
past of the system. The gravitational disturbances seem to be the triggering mechanism of the star-formation bursts %%@
that nowadays are observed in the knots of Mkn 1087. We propose the following scenario to explain the morphological %%@
features and chemical abundances present in the group:
\begin{enumerate}
\item Three independent galaxies form the group: Mkn 1087, KPG 103a and the north companion galaxy, this last one is %%@
less massive than the others. Mkn 1087 is a spiral galaxy nearly side-on. KPG 103a is probably another spiral galaxy.  
\item It is possible that Mkn 1087 and KPG 103a suffered a nearby encounter that formed the tidal features observed %%@
towards knots \#11 and \#12 from material stripped from Mkn 1087. We think that it also originated the tidal tail %%@
towards knot \#3, that then formed a tidal dwarf galaxy with an important old population (although some star-formation %%@
activity also occurred because it shows weak emission features). KPG 103a also developed some type of tidal features %%@
that now are difficult to detect; they are barely observed at the east of KPG 103a, although it also shows a diffuse %%@
plume towards the west. There are blue knots (possibly, star forming bursts) in the center of the galaxy, suggesting %%@
that perhaps KPG 103a is still suffering the effect of tidal perturbations.       
\item Some time later, the north companion object probably interacted with the north-east area of Mkn 1087. This %%@
encounter formed the tidal tails towards knot \#2 and the intense bridge of material towards knot \#1, perhaps %%@
triggering the starburst in both galaxies. This close encounter had to be intense and relatively recent, because the %%@
structures that originated them are clearly detected, and were made from material stripped from the main body of Mkn %%@
1087. Some star formation activity is also found in the tail to knot \#1. The very weak plume observed at the south of %%@
the north companion galaxy, that points towards knot \#1, reinforces this interacting scenario.
\end{enumerate}    

Both interacting processes were probably not timely separated, i.e, perhaps the interaction between Mkn 1087 and the %%@
north galaxy began before the interaction with KPG 103a ended. $N$-Body simulations [i.e. \citet{MH96,D98}] show that %%@
the elliptical galaxies found in most clusters and groups of galaxies have formed through the merger of several %%@
galaxies, although galactic cannibalism of smaller galaxies only accounts for a small fraction of the accreted mass. %%@
Hence, the probable future evolution of the system will be the merging of all galaxies and, maybe, an elliptical %%@
galaxy will be formed. It would be desirable to perform detailed studies of further LCBG in the Local Universe (as Mkn %%@
1087). This could be essential to understanding the origin and evolution of this kind of galaxies in the earlier %%@
Universe. Perhaps, galaxy merging and interactions played a more important role at large z.     

Finally, it should be very interesting to obtain \ion{H}{i} radio map of the system to analyze the distribution of the %%@
neutral gas presents in Mkn 1087 and its surroundings (in a similar way to the studies by \citet{HG96} or \citet{H01}, %%@
for example). This study will give a better understanding of the past and future evolution of the group galaxies, as %%@
well as it could allow to confirm definitely whether the tidal features observed in optical imagery are originated in %%@
the interactions between the different members of this group.    

\subsection{The non-catalogued cluster of galaxies.}

\begin{table}[t!]\centering
  \caption{\small{$B$-magnitude (not corrected by reddening) and $(B-R)$ color of some of the background galaxies (bg) %%@
found in the area surrounding Mkn 1087. The objects belonging to the non-catalogued cluster (cl) show redder colors.}}
  \label{table8}  
  \small
  \begin{tabular}{ccc}
   \noalign{\smallskip}
    \hline\hline
	\noalign{\smallskip}
   Object &   $m_B$  &  $(B-R)$ \\
    \hline
	\noalign{\smallskip}
    bg1    & 18.49$\pm$0.08  &  0.61$\pm$0.12 \\
	bg2    & 19.21$\pm$0.08  &  0.36$\pm$0.12 \\
	bg3    & 19.43$\pm$0.08  &  0.29$\pm$0.12 \\
	bg4    & 19.67$\pm$0.08  &  0.70$\pm$0.12 \\
	bg5    & 20.6$\pm$0.1    &  0.4$\pm$0.2 \\    
	bg6    & 21.5$\pm$0.2    &  0.4$\pm$0.3 \\
	bg7    & 21.6$\pm$0.2    &  0.1$\pm$0.3 \\
	bg8    & 21.8$\pm$0.2    &  0.6$\pm$0.3 \\
    \noalign{\smallskip}
    \hline
	\noalign{\smallskip}
	cl1    & 20.8$\pm$0.1    &  1.1$\pm$0.2 \\  
	cl2    & 21.8$\pm$0.2    &  1.0$\pm$0.3 \\
	cl3    & 21.9$\pm$0.2    &  1.8$\pm$0.3 \\
	cl4    & 22.0$\pm$0.2    &  2.3$\pm$0.3 \\
	cl5    & 22.5$\pm$0.2    &  2.7$\pm$0.3 \\
	cl6    & 23.2$\pm$0.3    &  2.8$\pm$0.5 \\ 
   \noalign{\smallskip}
   \hline\hline
  \end{tabular}  
 \end{table}

The cluster of galaxies found at the upper right corner of the image in Figure~\ref{figR} is not catalogued in the %%@
NASA/IPAC Extragalactic Database (NED). In Table~\ref{table8} we show the $B$-magnitude and $(B-R)$ color of some of %%@
the background galaxies (labeled with "bg") found in the deep $R$-image and observed in the $B$-image compared with %%@
the galaxies belonging to the cluster (labeled with "cl"). We performed a simple analysis of the surface brightness %%@
profiles of some of these background galaxies to confirm that they are not star-like objects. Despite the %%@
uncertainties due to the faintness of the galaxies with $m_B >$ 20, the objects of this non-catalogued cluster show %%@
redder colors that the rest of the background galaxies. 

It is possible to determine the redshift to the cluster using photometric redshift codes, for example, using the %%@
\citet{FS99} or HyperZ \citep{BMP00} codes, that find the redshift of each object using a fitting procedure that %%@
compares the observed magnitudes with the expected ones from standard spectral energy distributions (SED). However, to %%@
apply these methods to our cluster, it should be necessary to get the magnitude in, at least, another broad-band %%@
filter. Nevertheless, using the code that is been developed by Juncosa \& Guti\'errez [some previous results can be %%@
found in \citet{J04}], we estimate a redshift between 0.2 and 0.4 for it. The study of redshift clusters gives %%@
important clues about both galaxy and cluster formation and evolution.

\section{Conclusions}

We have used deep optical and near-infrared imagery and deep optical intermediate-resolution spectroscopy to analyze %%@
the morphology, colors, physical conditions, kinematics and chemical abundances of the galaxy Mkn 1087 and its %%@
surrounding objects. We have obtained the oxygen abundances for several star-forming bursts using empirical %%@
calibrations. We have obtained consistent age estimations for the knots observed in Mkn 1087 using different %%@
observational indicators and theoretical models. 

We have derived a Keplerian mass of $\sim 5.6 \times 10^{10}$ \Mo\ for the main body of Mkn 1087. We have detected a %%@
new companion object at the north, which has an oxygen abundance and N/O ratio lower that the rest of the knots, a %%@
Keplerian mass of $\sim 2.2 \times 10^8$ \Mo, an old underlying stellar population, and it is detached of the rotation %%@
pattern of the main galaxy. All these evidences suggest that the north companion object is an external dwarf galaxy. %%@
We suggest that some of the objects (\#1, \#2, \#3, \#11 and \#12) are TDG candidates. 

Mkn 1087 is not powered by an AGN but it could be classified as a LCBG because of its color, absolute magnitude and %%@
surface brightness. Wolf-Rayet features are not detected in the brightest bursts, perhaps due to aperture effects. %%@
Luminous knot \#7 hosts a strong starburst that should not be the consequence of a merger, more likely it is a complex %%@
of star-forming regions located in a spiral arm of Mkn 1087.

In conclusion, the complex geometry of the filamentary structure of Mkn 1087 and all the photometric, chemical and %%@
kinematical results can be explained assuming that it is in interaction with two external galaxies: 
\begin{enumerate}
\item the relatively bright KPG 103a, that could explain the bridges, the non-stellar objects located between both %%@
galaxies (\#11 and \#12), and the tidal dwarf galaxy \#3; 
\item and the new dwarf north companion, that could originate the tidal features towards knot \#2, the bridge between %%@
Mkn 1087 and the knot \#1, and produce the star formation triggering in the knots found in them. 
\end{enumerate}
We consider that Mkn 1087 and its surroundings should be classified as a group of galaxies in interaction. 

Finally, we have also serendipitously detected a non-cataloged cluster of galaxies with central position around %%@
$\alpha$ = 04h 49min 35sec, $\delta$ = +03$^\circ$ 21$\arcmin$ 04$\arcsec$ and probably at a redshift between 0.2 and %%@
0.4.

\begin{acknowledgements}
We are very grateful to the referee, Bill Keel, for his very useful comments. We would like to acknowledge Jos\'e A. %%@
Caballero for his macro for NIR reduction and V\'{\i}ctor S\'anchez-B\'ejar for his help with CST observations. We %%@
thank Vahram H. Chavushyan for helpful comments on the characteristics of AGNs. We are very grateful to Jorge %%@
Garc\'{\i}a-Rojas for his help in the derivation of the helium abundances and to Robert Juncosa for his valuable %%@
comments about photometric redshifts. M.R. acknowledges support from Mexican CONACYT project J37680-E. This research %%@
has made use of the NASA/IPAC Extragalactic Database (NED) which is operated by the Jet Propulsion Laboratory, %%@
California Institute of Technology, under contract with the National Aeronautics and Space Administration.
\end{acknowledgements}

%% The following command ends your manuscript. LaTeX will ignore any text
%% that appears after it.

\end{document}